\documentclass[journal]{IEEEtran}

\usepackage[utf8]{inputenc} 
\usepackage[T1]{fontenc}    
\usepackage{hyperref}       
\usepackage{url}            
\usepackage{booktabs}       
\usepackage{amsfonts}       
\usepackage{nicefrac}       
\usepackage{microtype}      
\usepackage{xcolor}         
\usepackage{graphicx}
\usepackage{subfigure}
\usepackage{amsmath, bm}
\usepackage{makecell}
\usepackage{multirow}
\usepackage{wrapfig}
\usepackage{wrapfig}
\usepackage{fancyvrb}

\VerbatimFootnotes

\makeatletter

\newcommand{\norm}[1]{\left\lVert#1\right\rVert}

\newcommand{\mean}[1]{\mathbb{E}\left[#1\right]}
\newcommand{\parn}[1]{\left(#1\right)}

\newlength{\NOTskip}

\title{StyleTTS: A Style-Based Generative Model for Natural and Diverse Text-to-Speech Synthesis}

\author{Yinghao~Aaron~Li, Cong~Han, and~Nima~Mesgarani}

\begin{document}

\maketitle

\begin{abstract}
Text-to-Speech (TTS) has recently seen great progress in synthesizing high-quality speech owing to the rapid development of parallel TTS systems. Yet producing speech with naturalistic prosodic variations, speaking styles, and emotional tones remains challenging. In addition, many existing parallel TTS models often struggle with identifying optimal monotonic alignments since speech and duration generation typically occur independently. Here, we propose StyleTTS, a style-based generative model for parallel TTS that can synthesize diverse speech with natural prosody from a reference speech utterance. Using our novel Transferable Monotonic Aligner (TMA) and duration-invariant data augmentation, StyleTTS significantly outperforms other baseline models on both single and multi-speaker datasets in subjective tests of speech naturalness and synthesized speaker similarity. Through self-supervised learning, StyleTTS can generate speech with the same emotional and prosodic tone as the reference speech without needing explicit labels for these categories. In addition, when trained with a large number of speakers, our model can perform zero-shot speaker adaption. The source code and audio samples can be found on our demo page at \url{https://styletts.github.io/}. 
\end{abstract}

\section{Introduction}

Text-to-speech (TTS), also known as speech synthesis, has made significant strides with the advent of deep learning, producing increasingly human-like synthetic speech \cite{kim2021conditional, jia2021png, tan2022naturalspeech}. However, synthesizing expressive speech that captures the full range of prosodic, temporal, and spectral features, also known as paralinguistic information, remains a challenge \cite{tan2021survey}. A single piece of text can be spoken in various ways, influenced by context, emotional tone, and a speaker's unique linguistic habits. Hence, TTS is fundamentally a one-to-many mapping problem requiring innovative approaches.

While several strategies have been proposed to address this, including methods based on variational inference \cite{kim2021conditional, hsu2018hierarchical, zhang2019learning, lee2020bidirectional}, flow-based modeling \cite{kim2021conditional, valle2020flowtron, kim2020glow}, controlling pitch, duration and energy \cite{valle2020mellotron, ren2021fastspeech}, and using external prosody encoder \cite{skerry2018towards, wang2018style, chen2021speech}, the production of synthesized speech still falls short of real human speech. In particular, accurately modeling and incorporating different speakers' speaking styles and emotional tones poses a significant challenge.

Many attempts have been made to integrate style information into TTS models \cite{skerry2018towards, wang2018style, sun2020fully, liu2021expressive}. These approaches are predominantly based on autoregressive models such as Tacotron. Non-autoregressive parallel TTS models, such as Fastspeech \cite{ren2019fastspeech} and Glow-TTS \cite{kim2020glow}, however, have several advantages over autoregressive models. These models generate speech in parallel, enabling fast speech synthesis, and they are also more robust to longer and out-of-distribution (OOD) utterances. Moreover, because phoneme duration, pitch, and energy are predicted independently from speech, models such as FastSpeech2 \cite{ren2021fastspeech} and FastPitch \cite{lancucki2021fastpitch} allow fully controllable speech synthesis. At the same time, these models have limitations. They predominantly concentrate on speech synthesis from a single target speaker and often achieve multi-speaker extensions by concatenating speaker embeddings with the encoder output. Models that explore speech styles
also incorporate styles by concatenating style vectors and phoneme embeddings as input to the
decoder. \cite{skerry2018towards, wang2018style, sun2020fully, liu2021expressive}. This approach may not capture the temporal variation of acoustic features in the target speech effectively. In contrast, the domain of style transfer introduces styles through conditional normalization methods like adaptive instance normalization (AdaIN) \cite{huang2017arbitrary}. This technique has proven effective in neural style transfer \cite{huang2018multimodal, liu2019few, choi2020stargan}, generative modeling \cite{karras2019style, karras2020analyzing, karras2021alias}, and neural image editing \cite{lee2020maskgan, zhu2020sean}. Application of these methods in speech synthesis has not been extensively explored yet, restricted primarily to voice conversion and speaker adaptation \cite{chou2019one, chen2021again, li21e_interspeech, chen2021adaspeech, min2021meta}.

The structure of parallel TTS models allows for the entire speech to be synthesized, presenting an opportunity to leverage the powerful AdaIN module for integrating generalized styles in diverse speech synthesis. Recent state-of-the-art models mostly employ the non-autoregressive parallel framework for TTS, but because they do not directly align the input text and speech like autoregressive models do, an external aligner such as the Montreal Forced Aligner \cite{mcauliffe2017montreal} that is pre-trained on a large dataset is often required. Since the external aligner is not trained on the TTS data and objectives, the alignments are not optimally suited for the TTS task. Although training internal aligners alleviates some generalization problems \cite{kim2021conditional, kim2020glow, miao2021efficienttts, elias2021parallel}, overfitting can occur as the aligners are trained on a smaller TTS dataset with only a mel-reconstruction loss.

Here, we introduce StyleTTS, a model that addresses the aforementioned challenges of incorporating diverse speaking styles and learning a reliable monotonic aligner. StyleTTS incorporates style-based generative modeling into a parallel TTS framework to enable natural and expressive speech synthesis. It leverages AdaIN to integrate style vectors derived from reference audio, capturing the full spectrum of a speaker's stylistic features. This allows our model to synthesize speech that emulates the prosodic patterns and emotional tones in the reference audio. With various reference audios, we can synthesize the same text in different speaking styles, effectively realizing the one-to-many mapping that many TTS systems find challenging. In addition, our model employs a novel Transferable Monotonic Aligner (TMA) to find the optimal text alignment, aided by a novel duration-invariant data augmentation scheme to produce naturalistic prosody robust to potentially suboptimal duration predictions. Our model’s design is robust against the generalization problems of external aligners and overfitting problems that can be caused by internal aligners.

This paper presents the following novel contributions: (i) the introduction of the Transferable Monotonic Aligner (TMA), a new transfer learning scheme that refines pre-trained text aligners for TTS tasks, (ii) a duration-invariant data augmentation method for improving prosody prediction, and (iii) a parallel TTS model that incorporates generalized speech styles for natural and expressive speech synthesis. Together, these contributions pave the way for advanced TTS technologies enhancing human-computer interactions.

\begin{figure*}[!ht]
\begin{center}
\vspace{-10 pt}
\centerline{\includegraphics[width=0.95\textwidth]{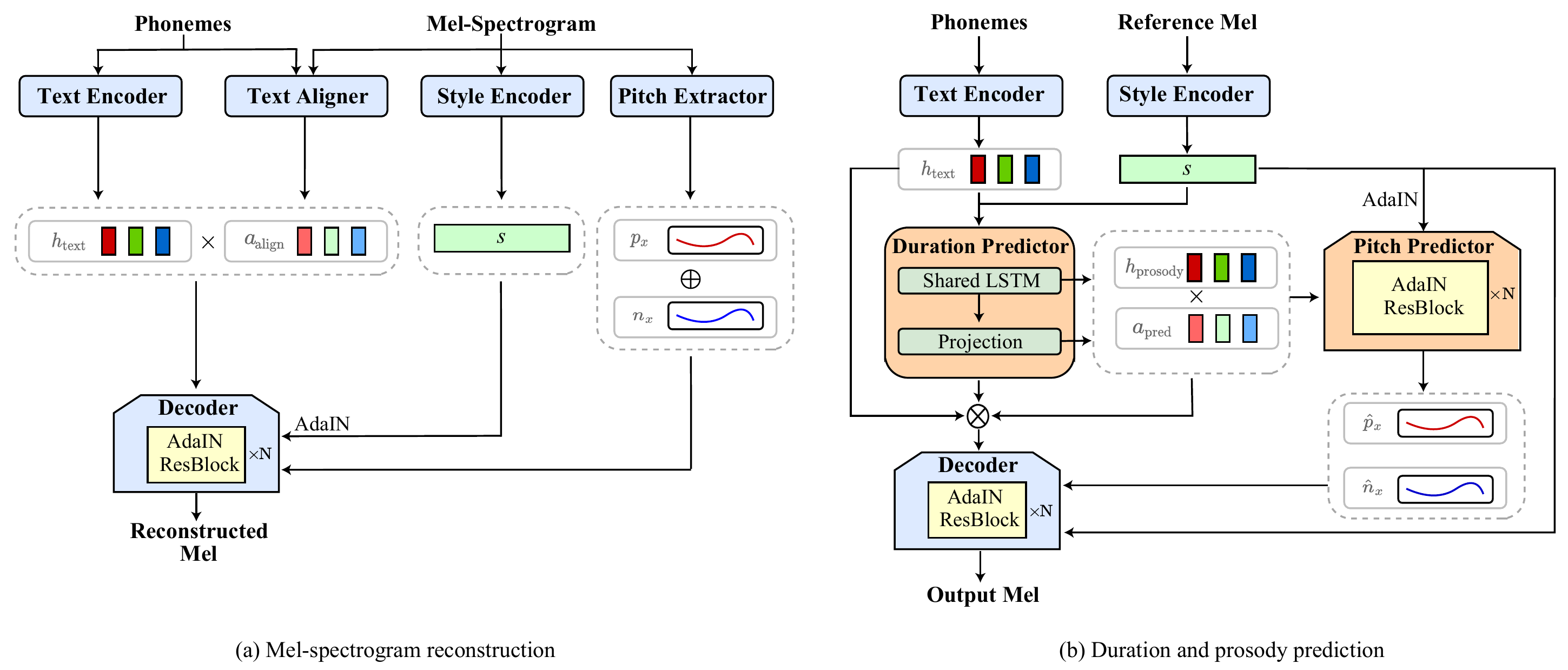}}
\caption{Training and inference schemes of StyleTTS. (a) Stage 1 of our training procedures where the decoder is trained to reconstruct input mel-spectrogram using pitch, energy, phonemes, alignment, and style vectors. (b) Stage 2 of training and inference procedures where pitch, energy, and alignment are predicted based on input text, and a style vector is extracted from a reference mel-spectrogram for synthesis. Parameters of modules in blue are fixed during this stage of training while those in orange are tuned. }
\label{fig:1}
\end{center}
\vspace{-15 pt}

\end{figure*}


\section{Methods}


\subsection{Proposed Framework}
\label{sec:2.1}
Given $\bm{t} \in \mathcal{T}$ the input phonemes and $\bm{x} \in \mathcal{X}$ an arbitrary reference mel-spectrogram, our goal is to train a system that generates the mel-spectrogram $\tilde{\bm{x}} \in \mathcal{X}$ that corresponds to the speech of $\bm{t}$ and reflects the generalized speech styles of $\bm{x}$. Generalized speech styles are defined as any characteristics in the reference audio $\bm{x}$ except the phonetic content  \cite{liu2021expressive}, including but not limited to prosodic pattern, lexical
stress, formants transition, speaking rate, and speaker identity. Our framework consists of eight modules that can be divided into three major categories: (i) speech generation modules that include the text encoder, style encoder, and decoder, (ii) TTS prediction modules that include the duration and prosody predictor, and (iii) utility modules only used during training that include the discriminator, text aligner, and pitch extractor. An overview of our framework is provided in Figure \ref{fig:1}. We detail each of the modules below. 

\textbf{Text Encoder. } The text encoder $T$ transforms the phonemes $\bm{t}$ into a hidden representation $\bm{h}_\text{text} = T(\bm{t})$. It consists of a 3-layer CNN followed by a bidirectional LSTM \cite{schuster1997bidirectional}. 

\textbf{Text Aligner. } The text aligner $A$ generates an alignment $\bm{d}_\text{align}$ between mel-spectrograms and phonemes. We train a text aligner $A$ alongside the decoder $G$ during the reconstruction phase. Modeled after the decoder of Tacorton 2 with attention, $A$ is initially trained on an automatic speech recognition (ASR) task using the LibriSpeech corpus \cite{panayotov2015librispeech} and then fine-tuned concurrently with our decoder. We call an aligner with this setup (pre-trained on large corpora and fine-tuned for TTS tasks) a \textbf{Transferable Monotonic Aligner} (TMA). 

\textbf{Style Encoder. } Given an input mel-spectrogram $\bm{x}$, our encoder derives a style vector $s = E(\bm{x})$. With various reference audios, $E$ can generate diverse style representations, allowing the decoder $G$ to create speech that mirrors the style $s$ of a reference audio $\bm{x}$. $E$ consists of four residual blocks \cite{he2016deep} followed by an averaging pooling layer along the time axis.

\textbf{Pitch Extractor. } As in FastPitch \cite{ren2021fastspeech}, we extract pitch F0 directly in Hertz without further processing, providing a more straightforward representation and allowing enhanced control of speech pitch. Instead of using the acoustic periodicity detection method \cite{boersma1993accurate} employed in FastPitch to estimate the ground truth pitch, we train a pitch extractor $F$ end-to-end with our decoder $G$ for a more accurate estimation. Our pitch extractor $F$ is a JDC network \cite{kum2019joint}, pre-trained on LibriSpeech with ground truth F0 estimated using YIN \cite{de2002yin}. This extractor is fine-tuned with the decoder to predict pitch $p_{\bm{x}} = F(\bm{x})$ for the reconstruction of $\bm{x}$.

\textbf{Decoder.} Our decoder $G$ is trained to reconstruct the input mel-spectrogram $\bm{x}$, represented by $\hat{\bm{x}} = G\parn{\bm{h}_\text{text} \cdot \bm{d}_\text{align}, \bm{s}, p_{\bm{x}}, n_{\bm{x}}}$. Here, $\bm{h}_\text{text} \cdot \bm{d}_\text{align}$ is aligned hidden representation of phonemes, $\bm{s} = E(\bm{s})$ is the style vector of $\bm{x}$, $p_{\bm{x}}$ is pitch contour of $\bm{x}$, and $ n_{\bm{x}}$ is energy (represented by the log norm) of $\bm{x}$ per frame. The decoder is comprised of seven residual blocks with AdaIN \cite{huang2017arbitrary}, defined as follows:
\begin{equation}
    \label{eq:adain}
    \text{AdaIN}(c, {s}) = L_\sigma({s}) \frac{c - \mu(c)}{\sigma(c)} + L_\mu({s}),
\end{equation} 
where $c$ is a single channel of the feature maps, $s$ is the style vector, $\mu(\cdot)$ and $\sigma(\cdot)$ denote the channel mean and standard deviation and $L_\sigma$ and $L_\mu$ are learned linear projections for computing the adaptive gain and bias using the style vector $s$. The use of AdaIN is one of the major differences between our model and other TTS models with style encoders such as \cite{wang2018style} and \cite{liu2021expressive}. The advantages of AdaIN for diverse speech synthesis are further discussed in Appendix \ref{sec:b.2}. 

To prevent dilution of import features, we concatenate the pitch $p_{\bm{x}}$, energy $n_{\bm{x}}$, and residual phoneme features $\bm{h}_\text{res}$ and deliver them to subsequent residual blocks after AdaIN. This process is further detailed in Table \label{tab:8}, and its effectiveness is discussed in Section \ref{sec:4.4}. 

\textbf{Discriminator. } We include a discriminator $D$ to facilitate training of our decoder for better sound quality \cite{kim2021conditional}. The discriminator shares the same architecture as our style encoder.

\textbf{Duration Predictor. } Our duration predictor consists of a 3-layer bidirectional LSTM $R$ with adaptive layer normalization (AdaLN) module followed by a linear projection $L$, where instance normalization is replaced by layer normalization in equation \ref{eq:adain}. We use AdaLN because $R$ takes discrete tokens similar to those in NLP applications, where layer normalization \cite{ba2016layer} is preferred. $R$ is shared with the prosody predictor $P$ through $\bm{h}_\text{prosody} = R\parn{\bm{h}_\text{text}}$  as input to $P$.

\textbf{Prosody Predictor. } Our prosody predictor $P$ predicts both the pitch $\hat{p}_{\bm{x}}$ and energy $\hat{n}_{\bm{x}}$ with given text and style vector. The aligned shared representation $\bm{h}_\text{prosody} \cdot \bm{a}$ is processed through a shared bidirectional LSTM layer to generate $\bm{h}_\text{prosody}$, which is then fed into two sets of three residual blocks with AdaIN and a linear projection layer, one for the pitch output and another for the energy output (see Appendix \ref{app:C} for details). 

\subsection{Training Objectives}
\label{sec:2.2.1}

Our model training process is divided into two stages to allow the integration of duration-invariant prosody data augmentation, a key contribution of our work. During the first stage, the model learns to reconstruct the mel-spectrogram from the text, pitch, energy, and style. The second stage fixes all modules except the duration and prosody predictors, which are trained to predict the duration, pitch, and energy from the given text. 

\subsubsection{First Stage Objectives}
\hfill

\textbf{Mel reconstruction. } Given a mel-spectrogram $\bm{x} \in \mathcal{X}$ and its corresponding text $\bm{t} \in \mathcal{T}$, we train our decoder under the $L_1$ reconstruction loss 
\begin{equation} \label{eq1}
\mathcal{L}_\text{mel} = \mathbb{E}_{\bm{x}, \bm{t}}\left[{\norm{\bm{x} - G\parn{\bm{h}_\text{text} \cdot \bm{a}_\text{align}, \bm{s}, p_{\bm{x}}, n_{\bm{x}}}}_1}\right] 
\end{equation}
where $\bm{h}_\text{text} = T(\bm{t})$ is the encoded phoneme representation, $\bm{a}_\text{align} = A(\bm{x}, \bm{t})$ is the attention alignment from the text aligner, $s = E(\bm{x})$ is the style vector of $\bm{x}$, $p_{\bm{x}} = F(\bm{x})$ is the pitch F0 of $\bm{x}$ and $n_{\bm{x}}$ is the energy of $\bm{x}$. For end-to-end (E2E) training with the decoder and the text aligner, we apply a novel 50\%-50\% strategy:  half the time, we use the raw attention output as the alignment, which allows gradient backpropagation through the text aligner; for the other half, we use the non-differentiable monotonic version of alignment through dynamic programming algorithms \cite{kim2020glow} to train the decoder for generating intelligible speech from monotonic hard alignment during inference. This innovative approach effectively fine-tunes the pre-trained text aligner to produce the optimal alignments for speech reconstruction, thus enhancing the sample quality of generated speech. The effectiveness of this strategy is analyzed in section \ref{sec:4.4}.

\textbf{TMA objectives. } We fine-tune our text aligner with the original sequence-to-sequence ASR objectives $\mathcal{L}_\text{s2s}$ to ensure that correct attention alignment is kept during the E2E training: 

\begin{equation} 
\label{eq3}
\mathcal{L}_\text{s2s} = \mathbb{E}_{\bm{x}, \bm{t}}\left[\sum\limits_{i=1}^N{\textbf{CE}(\bm{t}_i, \hat{\bm{t}}_i)}\right],
\end{equation}
where $N$ is the number of phonemes in $\bm{t}$, $\bm{t}_i$ is the $i$-th phoneme token of $\bm{t}$, $\hat{\bm{t}}_i$ is the $i$-th predicted phoneme token, and $\textbf{CE}(\cdot)$ is the cross-entropy loss function. 

Since this alignment is not necessarily monotonic, we use a simple L-1 loss $\mathcal{L}_\text{mono}$ that forces the soft attention alignment to be close to its non-differentiable monotonic version:
\begin{equation} \label{eq4}
\mathcal{L}_\text{mono} = \mathbb{E}_{\bm{x}, \bm{t}}\left[\norm{\bm{a}_\text{align} - \bm{a}_\text{hard}}_1\right],
\end{equation}
where $\bm{a}_\text{align} = A(\bm{x}, \bm{t})$ is the attention alignment and  $\bm{a}_\text{hard}$ is the monotonic hard alignment obtained through dynamic programming algorithms (see Appendix \ref{secb.1} for details).

\textbf{Adversarial objectives. } We employ two adversarial loss functions, the original cross-entropy loss function $\mathcal{L}_\text{adv}$ for adversarial training and the additional feature-matching loss \cite{salimans2016improved} $\mathcal{L}_\text{fm}$, to improve the sound quality of the reconstructed mel-spectrogram:

\begin{equation} \label{eq5}
\mathcal{L}_\text{adv} = \mathbb{E}_{\bm{x}, \bm{t}}\left[\log D(\bm{x}) + \log\parn{1 - D(\hat{\bm{x}})}\right],
\end{equation}

\begin{equation} \label{eq6}
\mathcal{L}_\text{fm} = \mathbb{E}_{\bm{x}, \bm{t}}\left[\sum\limits_{l=1}^{T} \frac{1}{N_l}\norm{D^l(\bm{x}) - D^l(\hat{\bm{x}})}_1\right],
\end{equation}
where $\hat{\bm{x}}$ is the reconstructed mel-spectrogram by $G$, $T$ is the total number of layers in $D$ and $D^l$ denotes the output feature map of $l$-th layer with $N_l$ number of features. The feature matching loss can be seen as a reconstruction loss of hidden layer features of real and generated speech as judged by the discriminator. 

\textbf{First stage full objectives. } Our full objective functions in the first stage can be summarized as follows with hyperparameters $\lambda_\text{s2s}$, $\lambda_\text{mono}$, $\lambda_\text{adv}$ and $\lambda_\text{fm}$:

\begin{equation} \label{eq7}
\begin{aligned}  
\min_{G, A, E, F, T} \max_{D} \text{    }&
 \mathcal{L}_\text{mel} + \lambda_\text{s2s}\mathcal{L}_\text{s2s} +  \lambda_\text{mono}\mathcal{L}_\text{mono}  \\ & 
+ \lambda_\text{adv}\mathcal{L}_\text{adv} + 
\lambda_\text{fm}\mathcal{L}_\text{fm}
\end{aligned}
\end{equation}

\subsubsection{Second Stage Objectives}
\label{sec:2.2.2}
\hfill

\textbf{Duration prediction. } We employ the $L$-1 loss to train our duration predictor

\begin{equation} \label{eq8}
\mathcal{L}_\text{dur} = \mathbb{E}_{\bm{d}}\left[\norm{\bm{d} - \bm{d}_\text{pred}}_1\right]
\end{equation}
where $\bm{d}$ is the ground truth duration obtained by summing $\bm{a}_{\text{align}}$ along the mel frame axis. $\bm{d}_\text{pred} = L(R(\bm{h}_\text{text}, \bm{s}))$ is the predicted duration under the style vector $\bm{s}$. 

\textbf{Prosody prediction. } We train our prosody predictor via a unique data augmentation scheme. Since the duration predictor is trained separately from other modules (using only $\mathcal{L}_\text{dur}$), the duration predictions it produces may not always be optimal or compatible with the prosody predictor. To make the prosody predictor more robust to these potentially suboptimal duration predictions, we augment the data to introduce prosody invariance over the duration.

More specifically, instead of using the ground truth alignment, pitch, and energy of the original mel-spectrogram, we first apply a 1-D bilinear interpolation to stretch or compress the mel-spectrogram in time to obtain the augmented sample $\bm{\tilde{x}}$. As a result, the speech speed changes, yet the pitch and energy curves remain consistent. Accordingly, the prosody predictor learns to maintain pitch and energy prediction invariance, regardless of the duration of speech. This approach helps mitigate issues with unnatural prosody when the predicted duration is incorrect.

We use $\mathcal{L}_{f_0}$ and $\mathcal{L}_n$, which are F0 and energy reconstruction loss, respectively:

\begin{equation} \label{eq9}
\mathcal{L}_\text{f0} = \mathbb{E}_{{p}_{\bm{\tilde{x}}}}\left[\norm{{p}_{\bm{\tilde{x}}} - P_p\parn{S\parn{\bm{h}_\text{text}, s} \cdot \bm{\tilde{a}}_\text{align}}}_1\right]
\end{equation}

\begin{equation} \label{eq10}
\mathcal{L}_{n} = \mathbb{E}_{\bm{\tilde{x}}}\left[\norm{{n}_{\bm{\tilde{x}}}- P_n\parn{S\parn{\bm{h}_\text{text}, s} \cdot \bm{\tilde{a}}_\text{align}}}_1\right]
\end{equation}
where ${p}_{\bm{\tilde{x}}}$, ${n}_{\bm{\tilde{x}}}$ and $\bm{\tilde{a}}_\text{align}$ are the pitch, energy and alignment of $\bm{\tilde{x}} \in \tilde{\mathcal{X}}$ the augmented dataset. $P_p$ denotes the pitch output from the prosody predictor, and $P_n$ denotes the energy output. 

\textbf{Decoder reconstruction. } Lastly, we aim to ensure that the predicted pitch and energy can be effectively used by the decoder. Given that the mel-spectrogram is stretched or compressed during data augmentation, using them as the ground truth may lead to unwanted artifacts in the predicted prosody. Instead, we train the prosody predictor to produce pitch and energy predictions that can effectively reconstruct the decoder's outputs

\begin{equation} \label{eq11}
\mathcal{L}_\text{de} = \mathbb{E}_{\bm{\tilde{x}}, \bm{t}}\left[{\norm{\hat{\bm{x}} - G\parn{\bm{h}_\text{text} \cdot \bm{\tilde{a}}_\text{align}, s, \hat{p}, \hat{n}}}_1}\right],
\end{equation}
where $\hat{\bm{x}} = G\parn{\bm{h}_\text{text} \cdot \bm{\tilde{a}}_\text{align}, s, \tilde{p}, \norm{\bm{\tilde{x}}}}$ is the reconstruction of $\bm{\tilde{x}} \in \tilde{\mathcal{X}}$, $\hat{p} = P_p\parn{S\parn{\bm{h}_\text{text}, s} \cdot \bm{\tilde{a}}_\text{align}}$ the predicted pitch and $\hat{n} = P_n\parn{S\parn{\bm{h}_\text{text}, s} \cdot \bm{\tilde{a}}_\text{align}}$ the predicted energy.

\textbf{Second stage full objectives. } Our full objective functions in the second stage can be summarized as follows with hyperparameters $\lambda_\text{dur}, \lambda_\text{f0}$, and $\lambda_{n}$:

\begin{equation} \label{eq12}
\begin{aligned}  
\min_{S, L, P} \text{    }
\mathcal{L}_\text{de} + \lambda_\text{dur}\mathcal{L}_\text{dur} + \lambda_\text{f0}\mathcal{L}_\text{f0} 
+ \lambda_{n}\mathcal{L}_{n}
\end{aligned}
\end{equation}

\section{Experiments}

\begin{figure}
\centering
\subfigure[An example speaker from ESD]{\label{fig:2a}\includegraphics[width=0.32\textwidth]{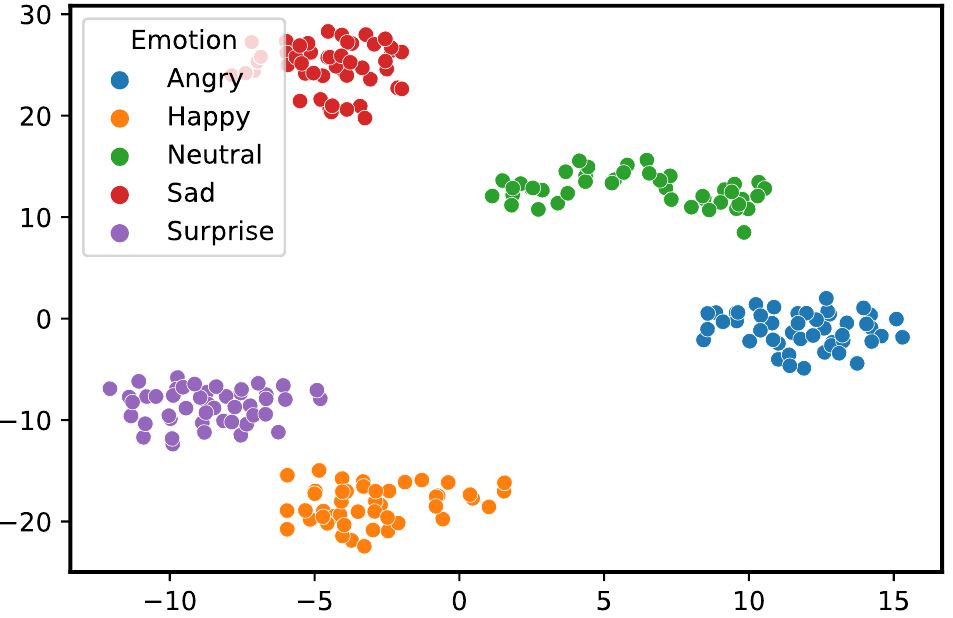}}
\subfigure[Same reference audios on LJ]{\label{fig:2b}\includegraphics[width=0.32\textwidth]{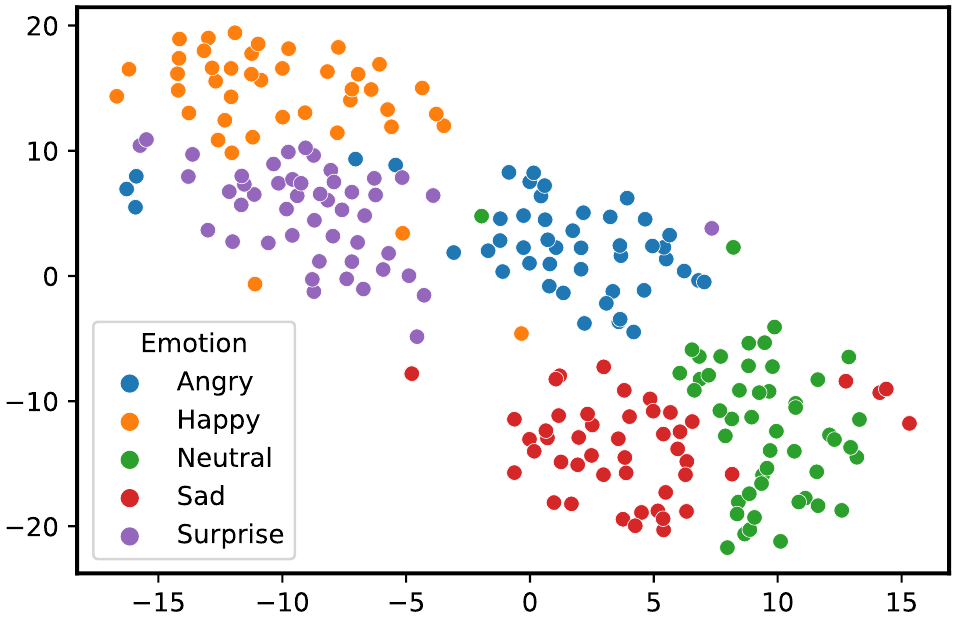}}
\subfigure[Unseen speakers from VCTK]{\label{fig:2c}\includegraphics[width=0.32\textwidth]{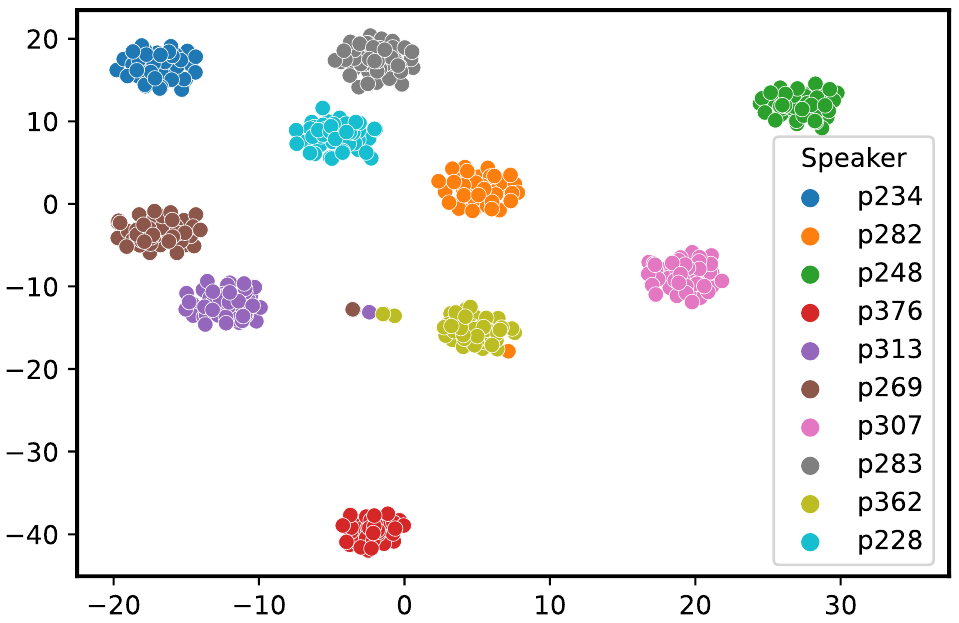}}
\caption{t-SNE visualization of style vectors. All styles are learned without explicit emotion or speaker labels. (a) Style vectors of reference audios in five different emotions of the speaker 0017 in ESD, computed by the multi-speaker model trained on ESD. (b) Style vectors of the same reference audios as in Fig. 2a, computed by the single-speaker model trained on the LJSpeech dataset. (c) Style vectors from the model trained on the LibriTTS data of 10 unseen speakers in the VCTK dataset. }
\vspace{-5 pt}

\end{figure}

\subsection{Datasets}
\label{dataset}
We conducted experiments on three datasets. We trained a single-speaker model on the LJSpeech dataset \cite{ljspeech17}. The LJSpeech dataset consists of 13,100
short audio clips with a total duration of approximately 24 hours. We used the same split as VITS where the training set contains 12,500 samples, the validation set 100 samples and the test set 500 samples. We also trained a multi-speaker model on the LibriTTS dataset \cite{zen2019libritts}. The LibriTTS {train-clean-460} subset consists of approximately 245 hours of audio from 1,151 speakers. We removed utterances with a duration longer than 30 seconds and shorter than one second. We randomly split the {train-clean-460} subset into a training (98\%), a validation (1\%), and a test (1\%) set and use the test set for evaluation following \cite{min2021meta}. We also used the VCTK \cite{yamagishi2019cstr} dataset to show that our model is capable of zero-shot speaker adaptation. We used the same training and test speaker split as in \cite{casanova2022yourtts} for VCTK, where 88 speakers were used for training and the rest 20 were used for testing. 

\begin{table}[!t]
\small
\caption{Comparison of evaluated MOS with 95\% confidence intervals (CI) on the LJ Speech dataset.}
\label{tab:1}
\centering
\begin{tabular}{ll}
\toprule
Model & MOS-N (CI) \\ 
\midrule
Ground Truth              & 4.32 ($\pm$ 0.04)\\
Tacotron 2 + HiFi-GAN     & 3.01 ($\pm$ 0.06)\\
FastSpeech 2 + HiFi-GAN   & 2.97 ($\pm$ 0.06)\\
VITS                      & 3.78 ($\pm$ 0.06)\\
StyleTTS + HiFi-GAN    &\textbf{4.01} ($\pm$ \textbf{0.05})\\
\bottomrule
\end{tabular}
\end{table}

In addition, we trained a multi-speaker model on the emotional speech dataset (ESD) \cite{zhou2021seen} to demonstrate the capacity to synthesize speech with diverse prosodic patterns. ESD consists of 10 Chinese and 10 English speakers reading the same 400 short sentences in five different emotions. We trained our model on 10 English speakers with all five emotions. 
We upsampled training audios to 24 kHz to match the LibriTTS dataset. We converted  text sequences into phoneme sequences using an open-source tool \footnote{\url{https://github.com/Kyubyong/g2p}}. We extracted mel-spectrograms with a FFT size of 2048, hop size of 300, and window length of 1200 in 80 mel bins using TorchAudio \cite{yang2021torchaudio}. 

\subsection{Training}
\label{train}
For both stages, we trained all models for 200 epochs using the AdamW optimizer \cite{loshchilov2018fixing} with $\beta_1 = 0, \beta_2 = 0.99$, weight decay $\lambda = 10^{-4}$, learning rate $\gamma = 10^{-4}$ and batch size of 64 samples. We set $\lambda_\text{s2s} = 0.2, \lambda_\text{adv} = 1, \lambda_\text{mono} = 5, \lambda_\text{fm} = 0.2, \lambda_\text{dur} = 1, \lambda_\text{f0} = 0.1$, and $\lambda_{n} = 1$. This setting of hyperparameters makes sure that all loss values are on the same scale and that the training is not sensitive to these hyperparameters. The scale factor ranges from 0.75 to 1.25 for data augmentation. We randomly divided the mel-spectrograms into segments of the shortest length in the batch. The training was conducted on a single NVIDIA A40 GPU.

\begin{figure}
\centering
\includegraphics[width=0.98\columnwidth]{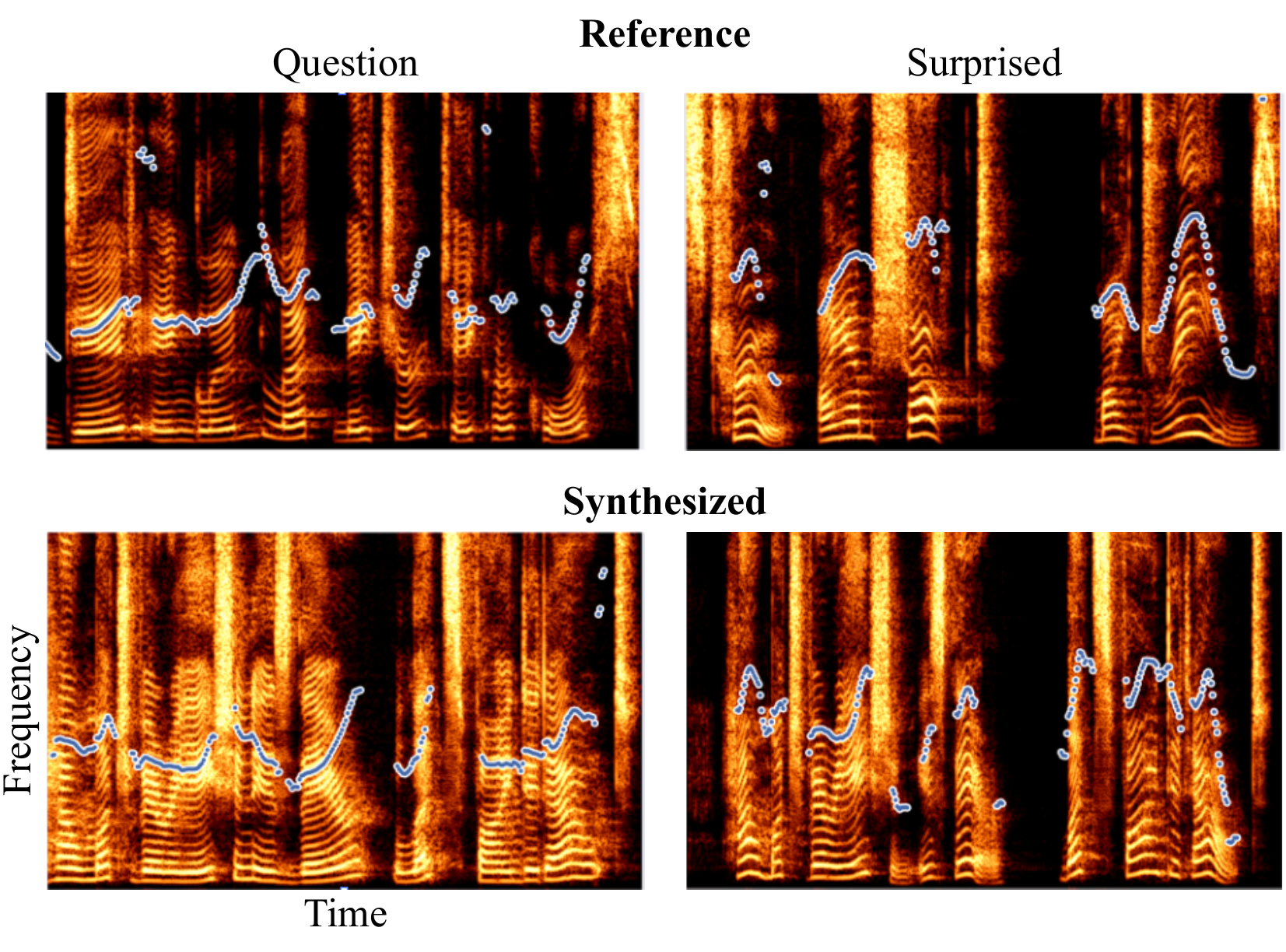}
\caption{Spectrograms of example reference audios and their corresponding generated speech reading ``How much variation is there? Let's find it out." from the single-speaker model trained on LJSpeech. The estimated pitch contour is shown as white dots. \textbf{Left top}: Reference audio of a question, ``Did England let nature take her course?". Note the pitch is mostly going up at the end of each word. \textbf{Left bottom}: Synthesized speech. The same pattern of pitch rising at the end of the words is present. \textbf{Right top}: Reference audio of surprised speech saying ``It's true! I am shocked! My dreams!". Note the pitch goes up first and then down for each word. \textbf{Right bottom}: Synthesized speech with the same pattern of the pitch going up and down for most of the words.}
\label{fig:3}
\end{figure}

\begin{figure*}[!t]
\centering
\subfigure[Pitch mean]{\label{fig:4a}\includegraphics[width=0.64\columnwidth]{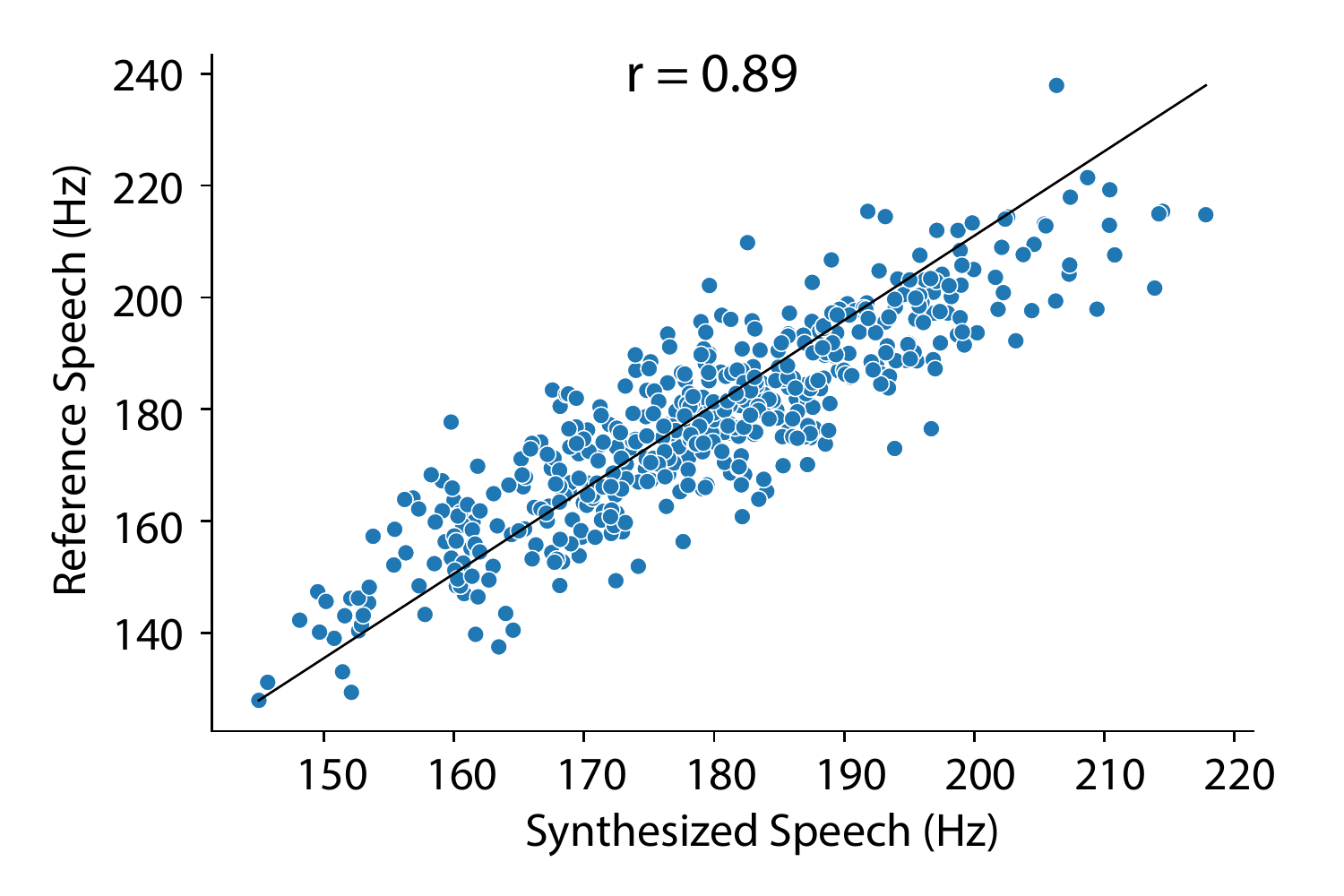}
}
\subfigure[Pitch standard deviation]{\label{fig:4b}\includegraphics[width=0.64\columnwidth]{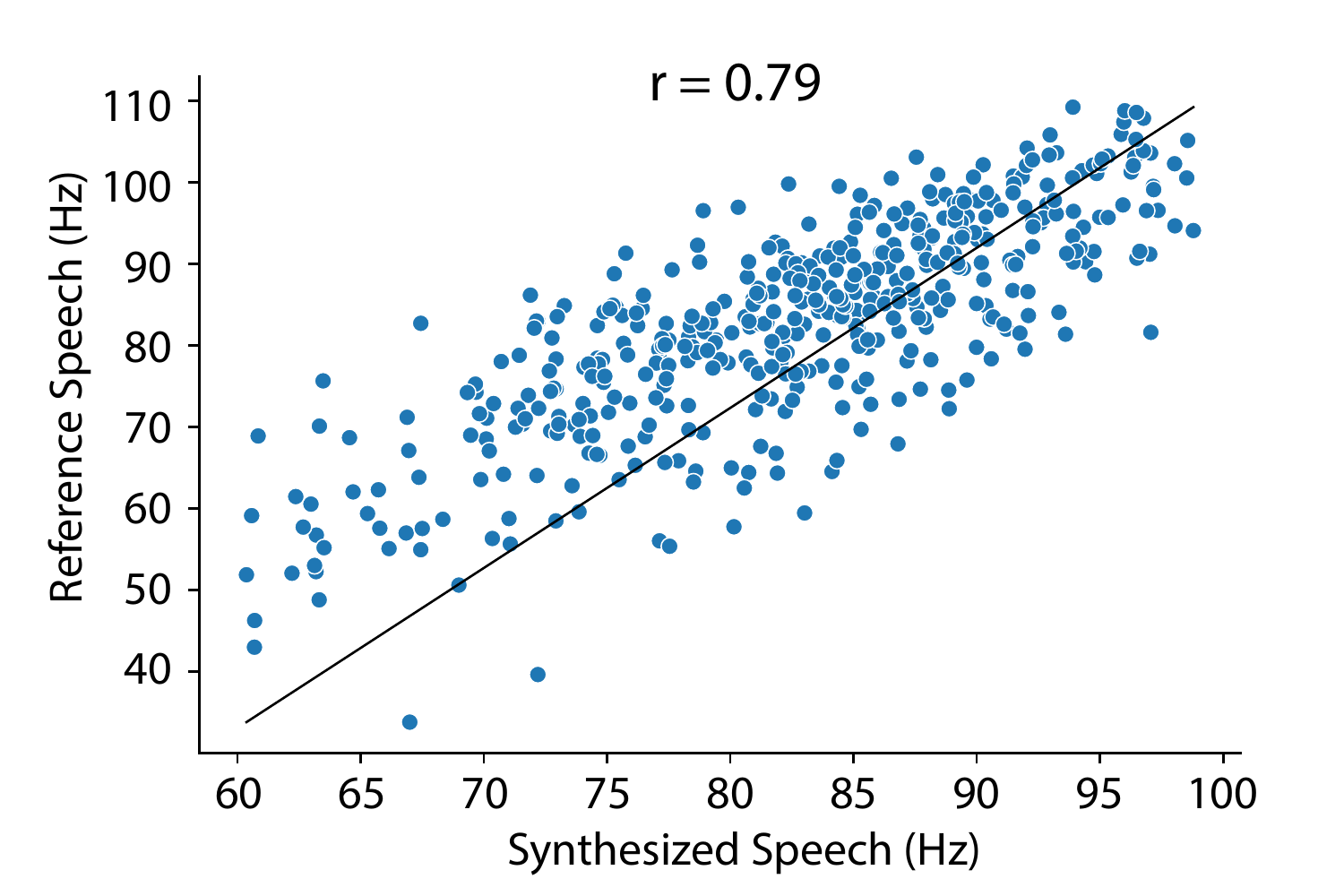}
}
\subfigure[Energy mean]{\label{fig:4c}\includegraphics[width=0.64\columnwidth]{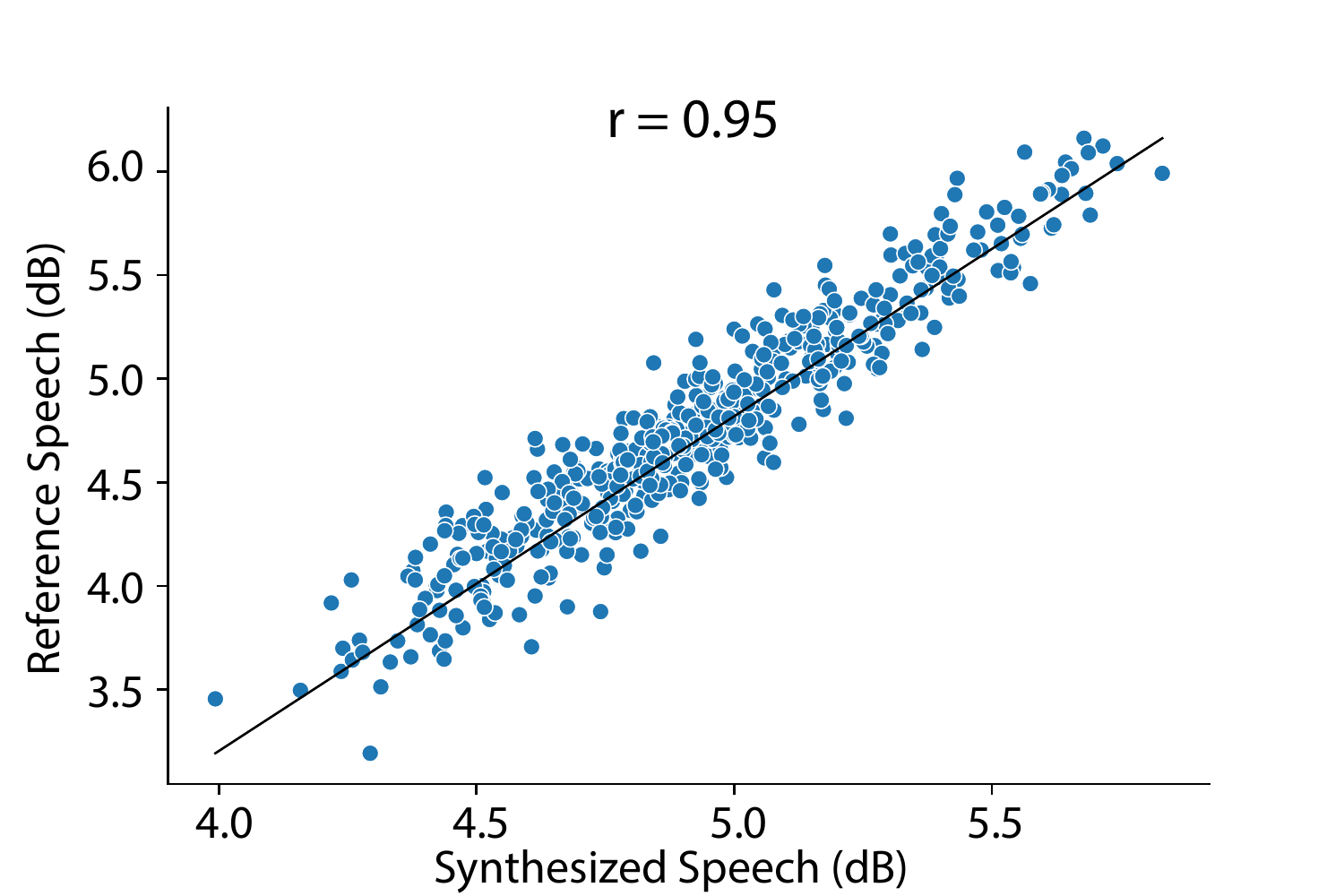}
}
\subfigure[Energy standard deviation]{\label{fig:4d}\includegraphics[width=0.64\columnwidth]{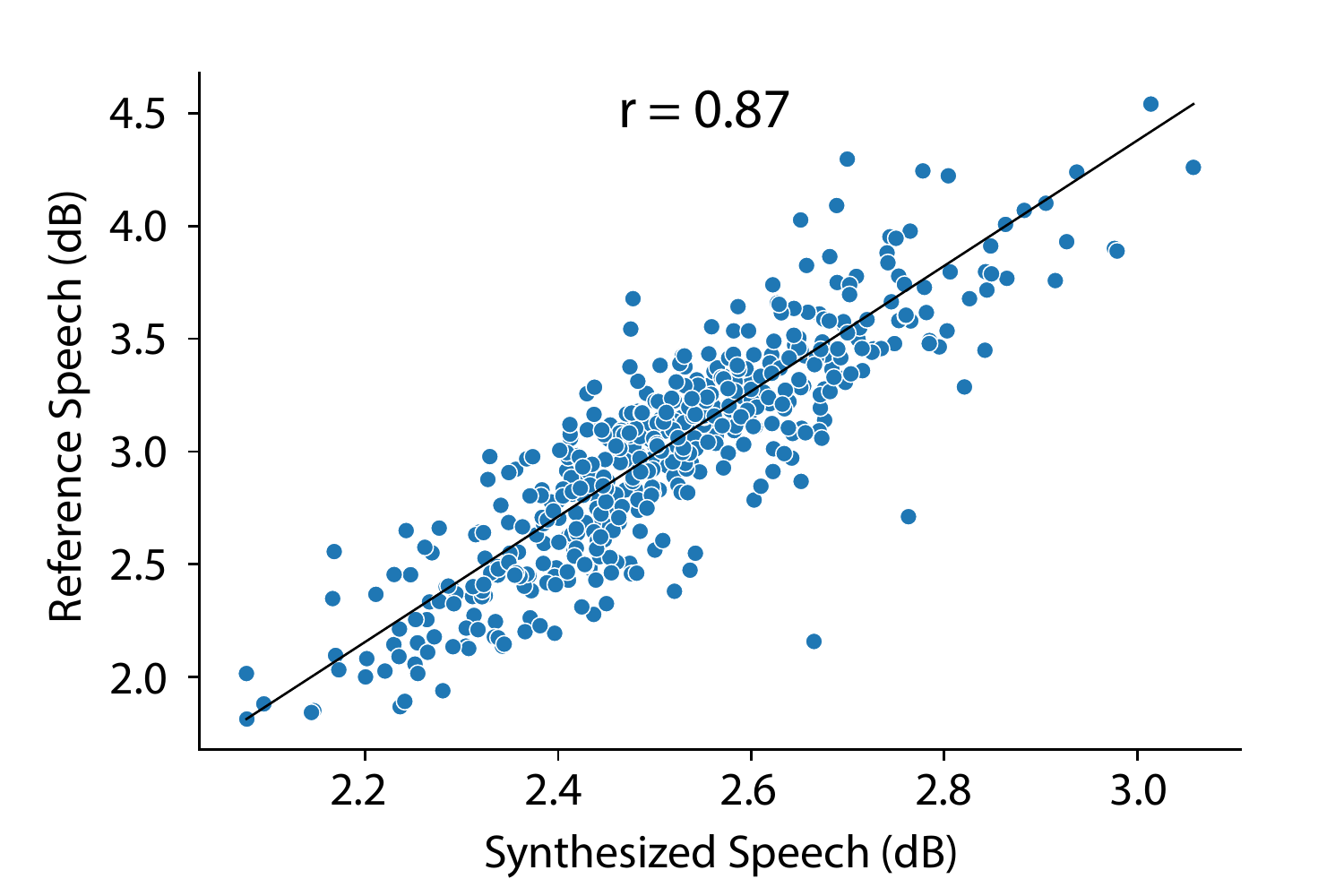}
}
\subfigure[Harmonics-to-noise ratio]{\label{fig:4e}\includegraphics[width=0.64\columnwidth]{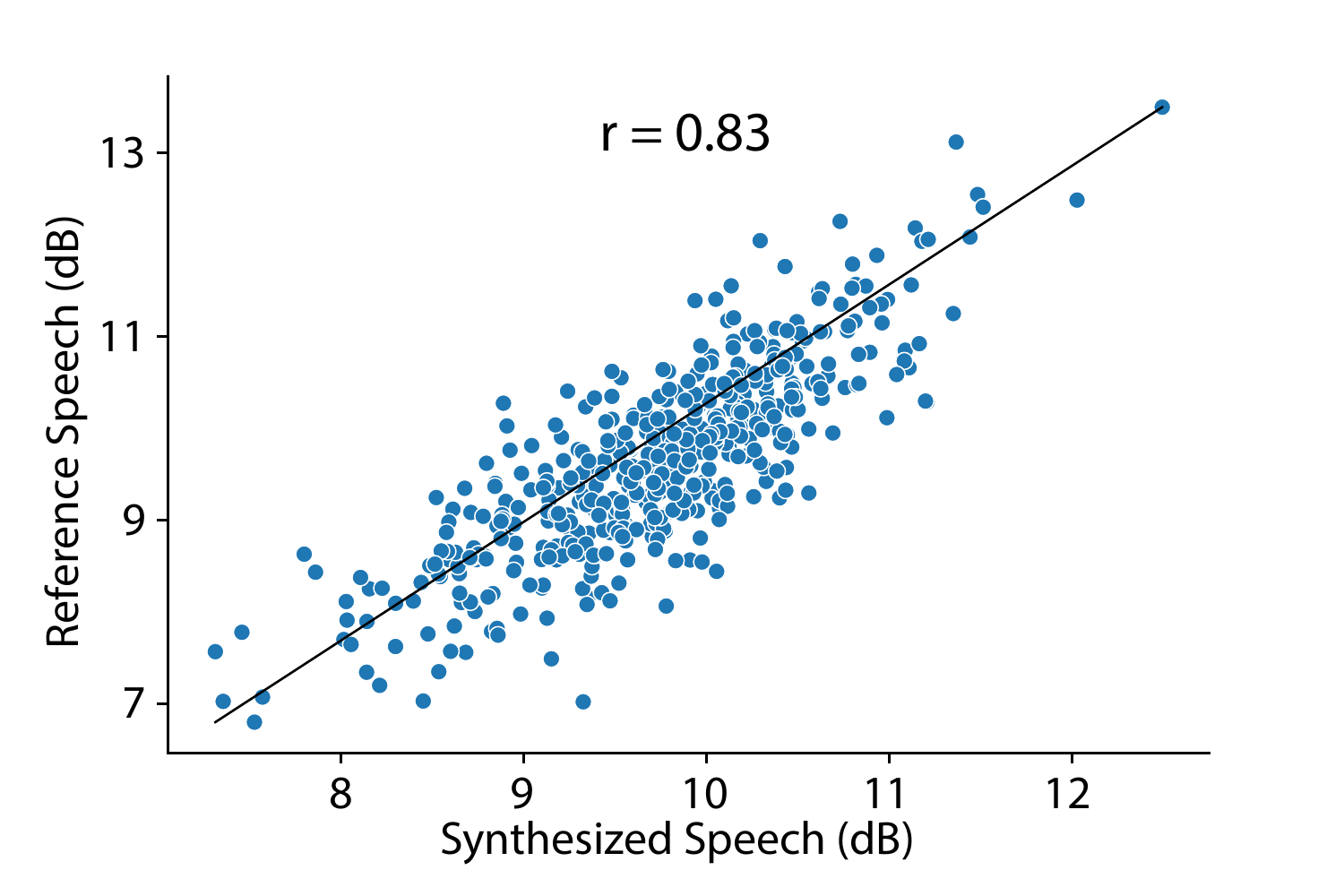}
}
\subfigure[Speaking rate]{\label{fig:4f}\includegraphics[width=0.64\columnwidth]{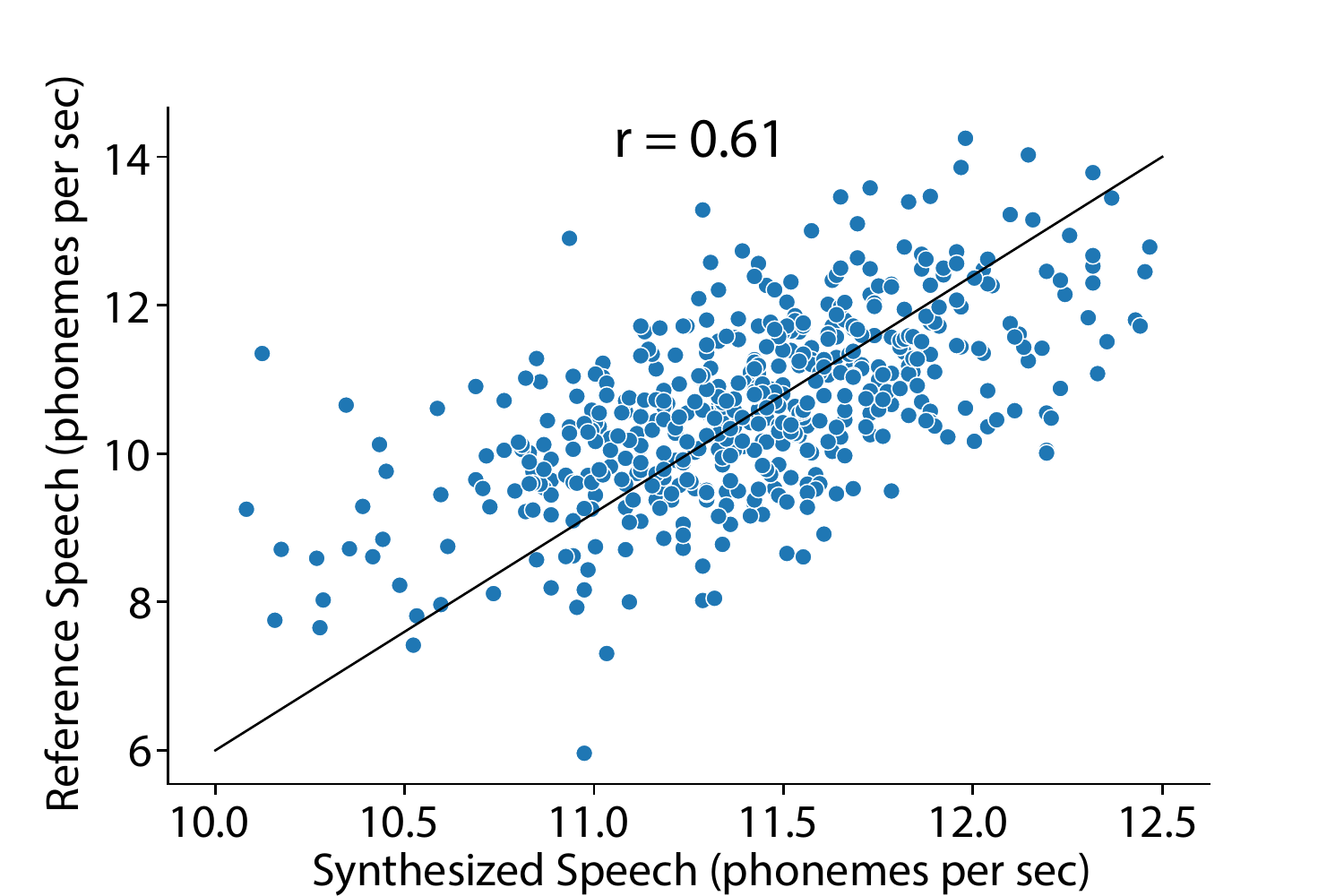}
}

\caption{Pearson correlation coefficients of six acoustic features associated with emotions between reference and synthesized speech on LJ Speech dataset.}
\label{fig:4}

\vspace{-5 pt}

\end{figure*}

\subsection{Evaluations}
\label{evalaution}

We performed two subjective evaluations:  mean opinion score of naturalness (MOS-N) to measure the naturalness of synthesized speech, and mean opinion score of similarity (MOS-S) to evaluate the similarity between synthesized speech and reference for the multi-speaker model. We recruited native English speakers located in the U.S. to participate in the evaluations on Amazon Mechanical Turk\footnote{We obtained approval for our protocol (number IRB-AAAR8655) from the Institutional Review Board (IRB) .}. In every experiment, we randomly selected 100 text samples from the test set. For each text, we synthesized speech using our model and the baseline models 
 and included the ground truth for comparison. The baseline models include Tacorton 2 \cite{shen2018natural}, FastSpeech 2 \cite{ren2021fastspeech}, 
 and VITS \cite{kim2021conditional}. For zero-shot speaker adapation experiments, we compared our model with StyleSpeech \cite{min2021meta} and YourTTS \cite{casanova2022yourtts}. All baseline models are pre-trained and publicly available  (see Appendix \ref{app:B} for details). The generated mel-spectrograms were converted into waveforms using the Hifi-GAN vocoder \cite{kong2020hifi} for all models. Each set of speech was rated by 10 raters on a scale from 1 to 5 with 0.5 point increments. For a fair comparison, we downsampled our synthesized audio into 22 kHz to match those from baseline models. We used random references when synthesizing speech for the single-speaker and zero-shot speaker adaption experiments. For multi-speaker models, because our training did not require speaker labels, for a fair comparison with other models that use explicit speaker embeddings during training, we averaged the style vectors computed using all samples in the training set from the same speaker as the reference style. 
 
 When evaluating each set, we randomly permuted the order of the models and instructed the subjects to listen and rate them without telling them the model labels. It is similar to multiple stimuli with hidden reference and anchor (MUSHRA), allowing the subjects to compare subtle differences among models. We used the ground truth as hidden attention checkers: raters were dropped from analysis if the MOS of the ground truth was not ranked top two among all the models.

\begin{table}[!t]
\small
  \caption{Comparison of evaluated MOS with 95\% confidence intervals (CI) on the LibriTTS dataset.}
  \label{tab:2}
  \centering
  \begin{tabular}{lll}
    \toprule
    Model & MOS-N (CI) & MOS-S (CI) \\ 
    \midrule
    Ground Truth              & 4.35 ($\pm$ 0.04) & 3.90 ($\pm$ 0.07)\\
    FastSpeech 2 + HiFi-GAN   & 3.00 ($\pm$ 0.06)  & 3.51 ($\pm$ 0.07)\\
    VITS                      & 3.62 ($\pm$ 0.06) & 3.70 ($\pm$ 0.07)\\
    StyleTTS + HiFi-GAN       & \textbf{4.03} ($\pm$ \textbf{0.05}) & \textbf{3.79} ($\pm$ \textbf{0.07})\\
    \bottomrule
  \end{tabular}
\end{table}

We also conducted objective evaluations using ASR metrics.  We evaluated the robustness of the models to different lengths of text input. We created four test sets with text length $L \textless 10$, $10 \leq L \textless 50$, $50 \leq L \textless 100$, and $100 \leq L$, respectively. Each set contains 100 texts sampled from the WSJ0 dataset \cite{garofolo1993csr}. We calculated the word error rate of the synthesized speech from both single-speaker and multi-speaker models using a pre-trained ASR model 
from ESPnet \cite{watanabe2018espnet}. To measure the inference speed, we computed the real-time factor (RFT), which denotes the time (in seconds) needed for the model to synthesize a one-second waveform. RFT was measured on a server with one NVIDIA 2080Ti GPU and a batch size of 1. In addition, we conducted the same analysis on the correlations of acoustic features associated with emotions between reference and synthesized speech using four multi-speaker models. Since there is no style in FastSpeech 2 and VITS, we used a pre-trained X-vector model \cite{snyder2018x} from Kaldi \cite{Povey_ASRU2011} to extract the speaker embedding as the reference vector.

\section{Results}

\begin{table}[!t]
\small
  \caption{Comparison of evaluated MOS with 95\% confidence intervals (CI) on the VCTK dataset for unseen speaker adaptation.}
  \label{tab:10}
  \centering
  \begin{tabular}{lll}
    \toprule
    Model & MOS-N (CI) & MOS-S (CI) \\ 
    \midrule
    Ground Truth              & 4.25 ($\pm$0.05) & 4.28 ($\pm$0.06)\\
    StyleTTS + HiFi-GAN     & \textbf{3.58} ($\pm$\textbf{0.06}) & \textbf{3.46} ($\pm$\textbf{0.07}) \\ 
    YourTTS  & 3.41 ($\pm$0.06) & 3.39 ($\pm$0.07)\\
    StyleSpeech + HiFi-GAN    & 2.16 ($\pm$0.05) & 2.43 ($\pm$0.06)\\
    \bottomrule
  \end{tabular}
\end{table}

\subsection{Model Performance}

Tables \ref{tab:1}, \ref{tab:2}, and \ref{tab:10} showcase the results of human subjective evaluations on the LJSpeech and LibriTTS datasets. When assessed for naturalness (MOS-N) and similarity (MOS-S), StyleTTS clearly outperforms other models, demonstrating its  superior performance under both single-speaker, multi-speaker, and zero-shot settings. Our models are more robust compared to other models (Table \ref{tab:5}), especially for long input texts. Since we do not use generative flows that require inverse Jacobian computation, our model is faster than VITS \cite{kim2021conditional}, even though it was not trained end-to-end like VITS (Table \ref{tab:6}).

We do note that our evaluation results differ from those reported in the baseline models, particularly for VITS. The VITS model has been reported to yield results very close to the ground truth \cite{kim2021conditional}. However, in our evaluation, VITS was found not to reach ground truth levels of performance. The primary factor leading to this discrepancy is the difference in evaluation methods. In VITS experiments, the traditional Mean Opinion Score (MOS) evaluation was used, where raters evaluated each module individually without any reference. The use of a reference point in our MUSHRA-like evaluation provides an anchor for rating, particularly the ground truth as the reference, which potentially lowers the scores of other models. A similar discrepancy has been reported in a very recent study that examines the effects of evaluation methods on the MOS results \cite{chiang2023we}, and our evaluation of VITS is comparable to other studies that have tried to reproduce it on both LJSpeech and LibriTTS datasets \cite{hayashi2021espnet2, lei2022glow, lim2022jets, liu2023diffvoice}. 

\begin{table*}[!t]
\small
\caption{Robustness evaluation on the LJSpeech and LibriTTS dataset. Word error rates (\%) are reported for different lengths of text (L).}
\label{tab:5}
\centering
\begin{tabular}{lccccr}
\toprule
 &  \multicolumn{4}{c}{WER ($\downarrow$)} \\ 
Model & $L \textless 10$ & $10 \textless L \textless 50$ & $50 \textless L \textless 100$ & $L \textgreater 100$ \\ 
\midrule
& \multicolumn{4}{c}{{\textit{Single-speaker models (on LJSpeech)}}} \\
Tacotron 2 + + HiFi-GAN & 17.22 & 12.61 & 16.95 & 46.33 \\
FastSpeech 2 + HiFi-GAN & 15.37 & 11.02 & 14.42 & 23.04 \\
VITS                    & 14.35 & 10.69 & 12.59 & 32.39 \\  
StyleTTS + HiFi-GAN     & \textbf{9.42}  & \textbf{7.44}  & \textbf{11.97} & \textbf{22.24} \\
\midrule
\midrule
& \multicolumn{4}{c}{{\textit{Multi-speaker models (on LibriTTS)}}} \\
FastSpeech 2 + HiFi-GAN  & \textbf{12.73} & 8.90  & 17.20 & 17.48 \\
VITS                     & 20.97 & 15.67 & 20.95 & 21.05\\  
StyleTTS + HiFi-GAN      & 17.35 & \textbf{8.26} &  \textbf{14.58} & \textbf{15.83} \\
\bottomrule
\end{tabular}
\end{table*}

\begin{table*}[!h]
\small
\caption{Comparison of Pearson correlation coefficients of acoustic features associated with emotions between reference and synthesized speech in multi-speaker experiments. Fastspeech 2 and VITS employ the X-vector as the reference.}
\label{tab:7}
\centering
\begin{tabular}{lccccccr}
\toprule
Model & \makecell{Pitch \\mean} & \makecell{Pitch \\standard \\deviation} & \makecell{Energy \\mean} & \makecell{Energy \\standard \\deviation} & \makecell{Harmonics-\\to-noise \\ratio}  
& Shimmer & Jitter \\
\midrule
FastSpeech 2    &0.95 &0.43 &0.23 & 0.51 & 0.81 
&0.81 &0.58\\
VITS                      &0.97  &0.32 &0.14 &0.5 &0.84 
&0.81 & 0.54\\
StyleTTS     &\textbf{0.99}  &\textbf{0.51} &\textbf{0.91} & \textbf{0.52}&\textbf{0.9} 
&\textbf{0.87} & \textbf{0.65}\\
\bottomrule
\end{tabular}
\end{table*}

\begin{table}[!ht]
\small
\caption{Real time factor (RTF) in second. }
\label{tab:6}
\centering
\begin{tabular}{lll}
\toprule
Model & RTF (s) \\ 
\midrule
Tacotron 2 + HiFi-GAN     &  0.0868\\
VITS                      &  0.0428\\
StyleTTS + HiFi-GAN       &  \textbf{0.0388}\\
\bottomrule
\end{tabular}
\end{table}

\subsection{Visualization of Style Vectors}

To verify that our model can learn meaningful style representations, we projected the style vectors extracted from reference audios into a 2-D plane for visualization using t-SNE \cite{van2008visualizing}. We selected 50 samples of each emotion from a single speaker in ESD and projected the style vectors of each audio into the 2-D space. It can be seen in Fig. \ref{fig:2a} that our style vector distinctively encodes the emotional tones of reference sentences even though the training does not use emotion labels. We also computed the style vectors using speech samples from the same speaker with our single-speaker model. This model is only trained on the LJSpeech dataset and therefore has never seen the selected speaker from ESD during training. Nevertheless, in Fig. \ref{fig:2b}, we see that our model can still clearly capture the emotional tones of the sentences, indicating that even when the reference audio is from a speaker different from the single speaker seen during training, it still can synthesize speech with the correct emotional tones. This shows that our model can implicitly extract emotions from an unlabeled dataset in a self-supervised manner. Lastly, we show projected style vectors from 10 unseen VCTK speakers each with 50 samples in Fig \ref{fig:2c}. Different speakers are perfectly separated from each other in the 2-D projection. This indicates that our model can learn speaker identities without explicit speaker labels and hence perform zero-shot speaker adaptation.

\subsection{Style-Enabled Diverse Speech Synthesis}
To show that the learned style vectors indeed enable diverse speech synthesis, we provide an example of synthesized speech with two different reference audios using our single-speaker model trained on the LJSpeech dataset in Figure \ref{fig:3}. It can be seen clearly that the synthesized speech captures various aspects of the reference speech, including the pitch, prosody, pauses, and formant transitions. To systematically quantify this effect, we drew six scatter plots showing the correlations between synthesized and reference speech in acoustic features traditionally used for speech emotion recognition (Figure \ref{fig:4}). 
The six features are pitch mean, pitch standard deviation, energy mean, energy standard deviation, harmonics-to-noise ratio, and speaking rate \cite{busso2013toward}. All six features demonstrate a significant correlation between the synthesized and reference speech ($p<0.001$) with the correlation coefficients all above 0.6. Our model also outperforms other models on multi-speaker datasets in acoustic feature correlations (Table \ref{tab:7}).
The results indicate that multiple aspects of the synthesized speech are matched to the reference, allowing flexible control over synthesized speech simply by selecting appropriate reference audios. Since our models also allow fully controllable pitch, energy, and duration, our approach is among the most flexible models in terms of controllability for speech synthesis.

\begin{table*}[!th]
\small
\centering
\caption{Ablation study for verifying the effectiveness of each proposed component. }
\label{tab:4}
    \begin{subtable}
    \centering
    \begin{tabular}{lcr}
    \toprule
    Model & CMOS \\ 
    \midrule
    StyleTTS                  & 0 \\
    w/ 100\% hard           & -- 0.26\\
    w/ 0\% hard          & -- 2.98 \\
    w/o $\mathcal{L}_\text{mono}$             & -- 0.10 \\
    w/o $\mathcal{L}_\text{s2s}$           & -- 2.48 \\
    \bottomrule
    \end{tabular}
    \end{subtable}
        \hfil
    \begin{subtable}
    \centering
    \begin{tabular}{lcr}
    \toprule
    Model & CMOS \\ 
    \midrule
    StyleTTS                  & 0 \\
    w/o pitch extractor             & -- 0.11\\
    w/o pre-trained aligner             & -- 0.39\\
    w/o augmentation            & -- 0.39\\
    w/o discriminator & -- 1.79 \\
    \bottomrule
    \end{tabular}
    \end{subtable}
    \hfil
    \begin{subtable}
    \centering
    \begin{tabular}{lcr}
    \toprule
    Model & CMOS \\ 
    \midrule
    StyleTTS                  & 0 \\
    w/o residual            & -- 0.30\\
    \text{AdaIN} $\rightarrow$ \text{AdaLN}             & -- 0.21\\
    \text{AdaIN} $\rightarrow$ \text{Concat.}             & -- 0.17\\
    \text{AdaIN} $\rightarrow$ \text{IN} & -- 0.03 \\
    \bottomrule
    \end{tabular}
    \end{subtable}
\end{table*}

\begin{table*}[!h]
\small
\caption{Comparison of Pearson correlation coefficients of acoustic features associated with emotions between reference and synthesized speech in ablation study.}
\label{tab:11}
\centering
\begin{tabular}{lccccccr}
\toprule
Model & \makecell{Pitch \\mean} & \makecell{Pitch \\standard \\deviation} & \makecell{Energy \\mean} & \makecell{Energy \\standard \\deviation} & \makecell{Harmonics-\\to-noise \\ratio}  
& Shimmer & Jitter \\
\midrule
Baseline    & \textbf{0.90} &\textbf{0.53} &\textbf{0.77} & 0.15 & \textbf{0.79} &\textbf{0.66} &0.64\\
{AdaIN} $\rightarrow$ {AadLN}    & 0.89 &0.52 &0.67 & \textbf{0.19} & 0.76 &0.53 &\textbf{0.66}\\
{AdaIN} $\rightarrow$ {Concat.}  & 0.36 & 0.16 &0.19 & -0.07 & 0.58 &0.36 & 0.40\\
w/o residual    & 0.88 &0.51 &0.68 & 0.11 & \textbf{0.79} &0.64 &0.60\\
\bottomrule
\end{tabular}
\end{table*}

\subsection{Ablation Study}
\label{sec:4.4}
We further conduct an ablation study to verify the effectiveness of each component in our model with experiments of subjective human evaluation. We instructed the subjects to compare our single-speaker model to those with one component ablated. We converted the ratings into comparative mean opinion scores (CMOS) by taking the difference between  the MOS of the baseline and ablated models. The results are shown in table \ref{tab:4}, and more details are in Appendix \ref{app:E}.

The leftmost table shows the results related to the proposed Transferable Monotonic Aligner (TMA) training. When training consists of 100\% hard alignments so that no gradient is back-propagated to the parameters of the text aligner (equivalent to using an external aligner such as in FastSpeech 2), the rated MOS is decreased by $-0.26$. This is due to the covariate shift between the pre-training data (LibriSpeech) and TTS data (LJ Speech). {An example of bad alignment of the pre-trained external aligner is shown in Figure \ref{fig:6}}. This shows that fine-tuning the aligner is effective in improving the quality of synthesized speech. However, when using 0\% hard alignment (100\%  soft attention alignment), the model gets overfitted to reconstruct speech with soft alignment and is unable to produce audible speech using hard alignment during inference ($-2.98$ CMOS). We also see that both TMA objectives (equations \ref{eq3} and \ref{eq4}) are important for high-quality speech synthesis. 

The table in the middle shows the effects of removing various training techniques and components. Using an external pitch extractor (such as acoustic-based methods) decreases MOS by $-0.11$. This is likely caused by the acoustic-based pitch extraction method sometimes failing to extract the correct F0 curve, and fine-tuning the pitch extractor along with the decoder makes the model learn better pitch representation (see Appendix \ref{pitch}). Without a pre-trained text aligner (such as VITS), the rated MOS is decreased by $-0.39$. This indicates that our transfer learning is helpful for mitigating overfitting problems when training internal aligners with a relatively small dataset. Removing our novel duration-invariant data augmentation also lowers the performance. Lastly, training without discriminators significantly affects the perceived sound quality. 

The rightmost table shows architecture changes by removing the residual features and replacing the AdaIN components in the decoder and predictor with instance normalization (IN), AdaLN, and simple feature concatenation (Concat). Their effects on style reflection are also shown in Table \ref{tab:11}. Removing the residual features in the decoder decreases both naturalness and correlations between synthesized and reference speech. Layer normalization is also worse than IN for both metrics. Concatenating styles in place of AdaIN dramatically decreases the correlations and lowers rated naturalness, confirming our observation that all previous methods that use concatenation to incorporate style information (\cite{kim2021conditional, kim2020glow, chen2020multispeech, skerry2018towards, wang2018style, sun2020fully, liu2021expressive}) are not as effective as AdaIN due to the lack of temporal modulations (see Appendix \ref{sec:b.2}). Lastly, we see that replacing AdaIN with IN does not significantly affect the rated naturalness, suggesting that the improved naturalness is not due to the introduction of styles but our novel technical improvements including TMA, data augmentation, use of IN, pitch extractor, and residual features. Nevertheless, styles enable diverse speech synthesis which models without styles cannot do.

\section{Conclusions}
\label{conclusions}
We introduced StyleTTS, a novel natural and diverse text-to-speech (TTS) synthesis approach. Our research takes a distinctive step forward in leveraging the strengths of parallel TTS systems with several novel constitutions that include a unique transferable monotonic aligner (TMA) training while integrating style information via AdaIN. We demonstrated that this method can effectively reflect stylistic features from reference audio. Moreover, the style vectors from our model encode a rich set of information present in the reference audio, including pitch, energy, speaking rates, formant transitions, and speaker identities. This allows easy control of the synthesized speech's prosodic patterns and emotional tones by choosing an appropriate reference style while benefiting from robust and fast speech synthesis of parallel TTS systems. Collectively, they enable natural speech synthesis with diverse speech styles that go beyond what was achieved in previous TTS systems. 

Our contribution lies not only in the theoretical underpinnings but also in its practical applicability. Our approach empowers various new applications, including movie dubbing, book narration, unsupervised speech emotion recognition, personalized speech generation, and any-to-any voice conversion (see Appendix \ref{app:D} and our follow-up work \cite{li2023stylettsvc} for more details). Our source code and pre-trained models are publicly available \footnote{\url{https://github.com/yl4579/StyleTTS}} to assist research in this area further.

\section{Acknowledgments}
We thank Gavin Mischler and Vinay Raghavan for their feedback. Funding was from the national institute of health (NIHNIDCD) and Marie-Josee and Henry R. Kravis.

\nocite{langley00}
\newpage
\bibliography{mybib}
\bibliographystyle{IEEEbib}



\newpage
\begin{appendices}
\section{Ablation Study Details}
\label{app:E}

In this section, we describe the detailed settings of each condition in Table \ref{tab:4} and provide more discussions of the results in Table \ref{tab:4} and Table \ref{tab:11}. 

\subsection{TMA-related} 
\label{secb.1}
\begin{figure*}[!th]
\includegraphics[width=\textwidth]{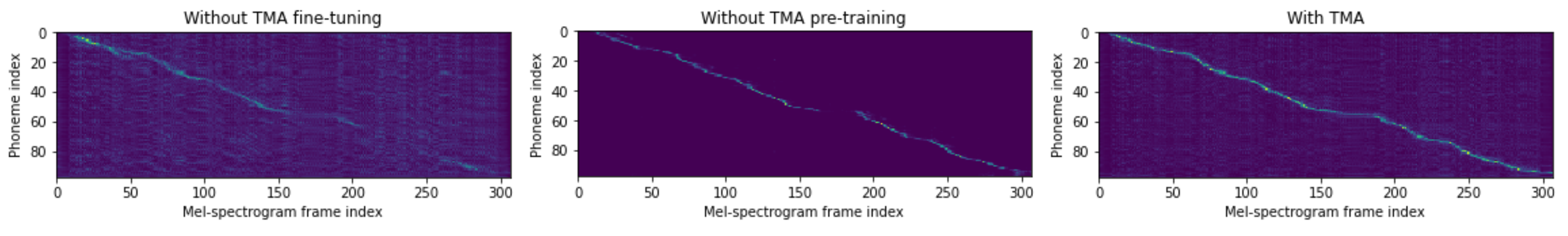}
\caption{An example showing text alignments under different conditions. \textbf{Left}: No TMA fine-tuning (100\% hard alignment such as FastSpeech). This is an example of a failed alignment. \textbf{Middle}: No  pre-trained text aligner (such as VITS). Note the gaps between alignments and clean attention (with no background noise), indicating some degrees of overfitting to the TTS speech dataset. \textbf{Right}: Full TMA fine-tuning. Note that TMA learns an alignment that is both continuous and monotonic compared to without fine-tuning and pre-training. }
\label{fig:6}
\end{figure*}

There are three Transferable Monotonic Aligner (TMA) related innovations in this work: the decoder is trained with hard monotonic alignment and soft attention in a 50\%-50\% manner and two TMA objectives functions. The 50\%-50\% training is motivated by the fact that the monotonic alignment search proposed in \cite{kim2020glow} is not differentiable, and the soft attention alignment does not necessarily provide correct alignments for duration prediction in the second stage of training. This 50\%-50\% split is arbitrary and can be changed to anything from 10\%-90\% to 90\%-10\%, depending on the dataset and the application. When the ratio is 100\%-0\%, it becomes the case where the external aligners are not fine-tuned like in most parallel TTS systems such as FastSpeech \cite{ren2019fastspeech}, while when the ratio is 0\%-100\%, it becomes the case we fine-tune the aligner with only soft attention such as in Cotatron \cite{park2020cotatron} for voice conversion applications. We find that training with external aligners (100\% hard, no fine-tuning) decreases the naturalness of the synthesized speech because bad alignments can happen due to covariate shifts between the training dataset (LibriSpeech) and testing dataset (LJSpeech) as in the case of Montreal Forced Aligner \cite{mcauliffe2017montreal}. One example is given in the leftmost figure in Figure \ref{fig:6}. On the other hand, if we only fine-tune the decoder with soft alignment, the decoder will overfit on the soft alignment and be unable to synthesize audible speech from hard alignment because the soft alignments are not either 0 or 1 and the precise numerical values of alignments are used by the decoder to generate speech. 

Another notable case is when we do not use a pre-trained text aligner such as in the case of VITS. This case makes MOS even lower than the case of no fine-tuning, suggesting that overfitting on a smaller dataset can be more detrimental than failure in generalization on the TTS dataset for some samples. The figure in the middle in Fig. \ref{fig:6} shows an alignment with gaps and no background noises. This indicates overfitting of the text aligner to the smaller dataset for the mel-spectrogram reconstruction objective. However, since our goal is to synthesize the speech from predicted alignment, overfitting to speech reconstruction can be harmful to natural speech synthesis during inference. 

In addition to the 50\%-50\% training, we also introduced two TMA objectives $\mathcal{L}_{s2s}$ and $\mathcal{L}_{mono}$. This is motivated by the fact that $\mathcal{L}_{s2s}$ learns correct alignments for S2S-ASR but not necessarily monotonic while non-differentiable monotonic alignments obtained through dynamic programming algorithms proposed in \cite{kim2020glow} do not necessarily produce correct alignments. By combining $\mathcal{L}_{s2s}$ and $\mathcal{L}_{mono}$, we can learn an aligner that produces both correct and monotonic alignments. 

\subsection{AdaIN, AdaLN, and Concatenation}
\label{sec:b.2}
As shown in Table \ref{tab:4} and Table \ref{tab:11}, AdaIN outperforms AdaLN and simple concatenation for both naturalness and style reflection. Here we describe our intuitions behind these results. 

\textbf{Concatenation vs. AdaIN. } When we concatenate the style vector to each frame of the encoded phonetic representations, we create a representation $\bm{h}_{\text{style}} = \begin{bmatrix}
\bm{h}_{\text{text}} \\ \text{---} \\ \bm{s}
\end{bmatrix}$. When the $\bm{h}_{\text{style}}$ is passed to the next convolution layer whose parameter is $W$, we get 

\begin{equation}
\begin{aligned}  
    \bm{h}_{\text{style}} \cdot W & = \begin{bmatrix}\bm{h}_{\text{text}}  \\ $---$ \\   \bm{s}\end{bmatrix} \cdot \begin{bmatrix}W_{\text{text}}  |   W_{\text{style}}\end{bmatrix} \\ & = \bm{h}_{\text{text}} \cdot W_{\text{text}} + \bm{s} \cdot W_{\text{style}}\\ & =  \bm{h}_{\text{text}} \cdot W_{\text{text}} +  \text{Concat }(\bm{h}_{\text{text}} , \bm{s})
\end{aligned}
\end{equation}

\noindent where $W_{\text{text}}$ and $W_{\text{style}}$ are block matrix notation of the corresponding weights for $\bm{h}_{\text{style}}$ and $\bm{s}$ and $\text{Concat }(\bm{h}_{\text{text}} , \bm{s}) = \bm{s} \cdot W_{\text{style}}$ denotes the concatenation operation as a function of input $\bm{h}_{\text{text}}$ and style vector $\bm{s}$. This \text{Concat }$(\bm{x}, s)$ function is almost like AdaIN in equation \ref{eq:adain} where $L_{\mu}(s) = W_{\text{style}}$ except we do not have the temporal modulation term $L_\sigma(s)$. The modulation term is very important in style transfer, and some works argue that modulation alone is enough for diverse style representations \cite{karras2020analyzing, an2021artflow}. In contrast, concatenation only provides the addition term ($L_\mu$) but no modulation term ($L_\sigma$). Intuitively, the modulation term can determine the variance of the pitch and energy, for example, and therefore without such a term, correlations for pitch and energy standard deviation  are much lower than AdaIN and AdaLN as shown in Table \ref{tab:11}.

\textbf{AdaLN vs. AdaIN. } Generative models for speech synthesis learn to generate mel-spectrograms, which is essentially a 1-D feature map with 80 channels. Each channel in the mel-spectrogram represents a single frequency range. When we apply AdaIN, we learn a distribution with a style-specific mean and variance for \textit{each channel}, compared to AdaLN, where a single mean and variance are learned for the \textit{entire feature map}. This inherent difference between feature distributions makes AdaIN more expressive in terms of style reflection than AdaLN.

\subsection{Pitch Extractor}
\label{pitch}
Acoustic methods for pitch estimation sometimes fail because of the presence of non-stationary speech intervals and sensitivity of hyper-parameters as discussed in the original papers that propose these methods \cite{boersma1993accurate, de2002yin}. A neural network trained with ground truth from these methods, however, can leverage the problems of failed pitch estimation because the failed pitch estimation can be regarded as noises in the training set, so it does not affect the generalization of the pitch extractor network. Moreover, since the pitch extractor is fine-tuned along with the decoder, there is no ground truth for the pitch beside the sole objective that the decoder needs to use extracted pitch information to reconstruct the speech. This fine-tuning allows better pitch representations beyond the original F0 in Hertz, but it also allows flexible pitch control as we can still recognize the pitch curves and edit them later when needed during inference. 

\section{Subjective Evaluation Details}
\label{app:B}

We used the publicly available pre-trained models as baselines for comparison. For the single-speaker experiment on the LJSpeech dataset, we used pre-trained Tacotron2\footnote{The model was \verb|kan-bayashi/ljspeech_tacotron2| from ESPNet}, Fastspeech2 \footnote{The model was \verb|kan-bayashi/ljspeech_fastspeech2| from ESPNet}, HiFiGAN \footnote{The model was \verb|parallel_wavegan/ljspeech_hifigan.v1| from ESPNet} from ESPnet. We used VITS \footnote{The implementation can be found at  \url{https://github.com/jaywalnut310/vits}} and YourTTS \footnote{The implementation can be found at  \url{https://github.com/Edresson/YourTTS}} from the official implementation. We randomly selected 100 text samples from the test set to synthesize the speech. Since audios from our model were synthesized using Hifi-GAN trained with audios sampled at 24 kHz, for a fair comparison, we resampled all the audios into 22 kHz and then normalized their amplitude. We used the pre-trained model for StyleSpeech \cite{min2021meta} \footnote{The implementation can be found at \url{https://github.com/jaywalnut310/vits}} from a public repository in GitHub for comparison of zero-shot speaker adaptation in Appendix \ref{app:D}. We did not use the official implementation because the vocoder used was MelGAN sampled at 16 kHz while the implementation we employed uses Hifi-GAN sampled at 22 kHz, which is comparable to other models. 

To reduce the listening fatigue, we randomly divided these 100 sets of audios into 5 batches \footnote{The survey (batch 1) can be found at \url{https://survey.alchemer.com/s3/6696322/LJ100-B1}} with each batch containing 20 sets of audios for comparison. We launched the 5 batches sequentially on Amazon Mechanical Turk (AMT) \footnote{\url{https://www.mturk.com/}}. We required participating subjects to be native English speakers located in the United States. For each batch, we made sure that we had collected completed responses from at least 10 self-reported native speakers whose IP addresses were within the United States and residential (i.e., not VPN or proxies). We used the average score that a subject rated on ground truth audios to check whether this subject carefully finished the survey as the subjects did not know which audio was the ground truth. We then disqualified and dropped all ratings from the subjects whose average ground truth score was not ranked top two among all the models. Finally, 46 out of 50 subjects were qualified for this experiment. 

\begin{table*}[!htbp]
\small
\caption{Decoder architecture. $T$ represents the input length of the mel-spectrogram, $p$ is the input F0, $n$ is the input energy, and $s$ is the style code. $\tilde{n}$ and $\tilde{p}$ are the processed pitch and energy, and $h_\text{res}$ is the output of the phoneme residual sub-module. }
\label{tab:8}
\centering
\begin{tabular}{ccccc}
\hline
Submodule                          & External Input                    & Layer          & Norm  & Output Shape      \\ \hline
\multirow{3}{*}{F0 processing}     & $p$                 & Input F0 $p$     & -     & 1$\times T$               \\
                                   & -                        & ResBlk         & -     & 64$\times T$              \\
                                   & -                        & Conv 1$\times$1       & IN    & 1$\times T$               \\ \hline
\multirow{3}{*}{Energy processing} & $n$                   & Input energy $n$ & -     & 1$\times T$               \\
                                   & -                        & ResBlk         & -     & 64$\times T$              \\
                                   & -                        & Conv 1$\times$1       & IN    & 1$\times T$               \\ \hline
\multirow{2}{*}{Phoneme residual}  & $h_{\text{text}}$                        & Input $h_{\text{text}}$         & -     & 512$\times T$             \\
                                   & -                        & Conv 1$\times$1       & IN    & 64$\times T$              \\ \hline
\multirow{3}{*}{IN ResBlks}        &  $\tilde{p}$, $\tilde{n}$,   $h_{\text{res}}$          & Concat         & -     & (512 + 2)$\times T$       \\
                                   & -                        & ResBlk         & IN    & 1024$\times T$            \\
                                   & -                        & ResBlk         & IN    & 1024$\times T$            \\ \hline
\multirow{9}{*}{AdaIN ResBlks}     &  $\tilde{p}$, $\tilde{n}$,  $h_{\text{res}}$  & Concat         & -     & (1024 + 2 + 64)$\times T$ \\
                                   & $s$                         & ResBlk         & AdaIN & 1024$\times T$            \\
                                   &   $\tilde{p}$, $\tilde{n}$,   $h_{\text{res}}$  & Concat         & -      & (1024 + 2 + 64)$\times T$ \\
                                   & $s$                        & ResBlk         & AdaIN & 1024$\times T$            \\
                                   &  $\tilde{p}$, $\tilde{n}$,   $h_{\text{res}}$    & Concat         & -     & (1024 + 2 + 64)$\times T$ \\
                                   & $s$                        & ResBlk         & AdaIN & 512$\times T$             \\
                                   & $s$                         & ResBlk         & AdaIN & 512$\times T$             \\
                                   & $s$                        & ResBlk         & AdaIN & 512$\times T$             \\
                                   & -                        & Conv 1$\times$1       & -     & 80$\times T$              \\ \hline
\end{tabular}
\end{table*}

In the multi-speaker experiments, we used pre-trained Fastspeech2\footnote{The model was \verb|kan-bayashi/libritts_xvector_conformer| \\ \verb|_fastspeech2| from ESPNet} , VITS \footnote{The model was \verb|kan-bayashi/libritts_xvector_vits| from ESPNet} , and HiFiGAN \footnote{The model was \verb|parallel_wavegan/libritts_hifigan.v1| from ESPNet} from ESPnet. We used pre-trained VITS from ESPnet instead of the official repository because we need the model to be trained on the LibriTTS dataset; however, the official models were trained on the LJSpeech or VCTK dataset. 

Similar to the single-talker experiment, we launched 5 batches \footnote{The survey (batch 1) can be found at \url{https://survey.alchemer.com/s3/6705095/LibriTTS-seen100-B1}} on AMT when we tested the multi-talker models on the LibriTTS dataset. 48 out of 58 subjects were qualified. We launched 3 batches \footnote{The survey (batch 1) can be found at \url{https://survey.alchemer.com/s3/6706053/zero-shot-B1}} with batch sizes 33, 33, 34, respectively, when we tested the multi-talker models on the VCTK dataset. 28 out of 30 subjects were qualified.

\section{Zero-Shot Voice Conversion}

\label{app:D}

\begin{figure*}[!htbp]
\centering
\subfigure[Source audio]{\label{fig:5a}\includegraphics[width=0.31\textwidth]{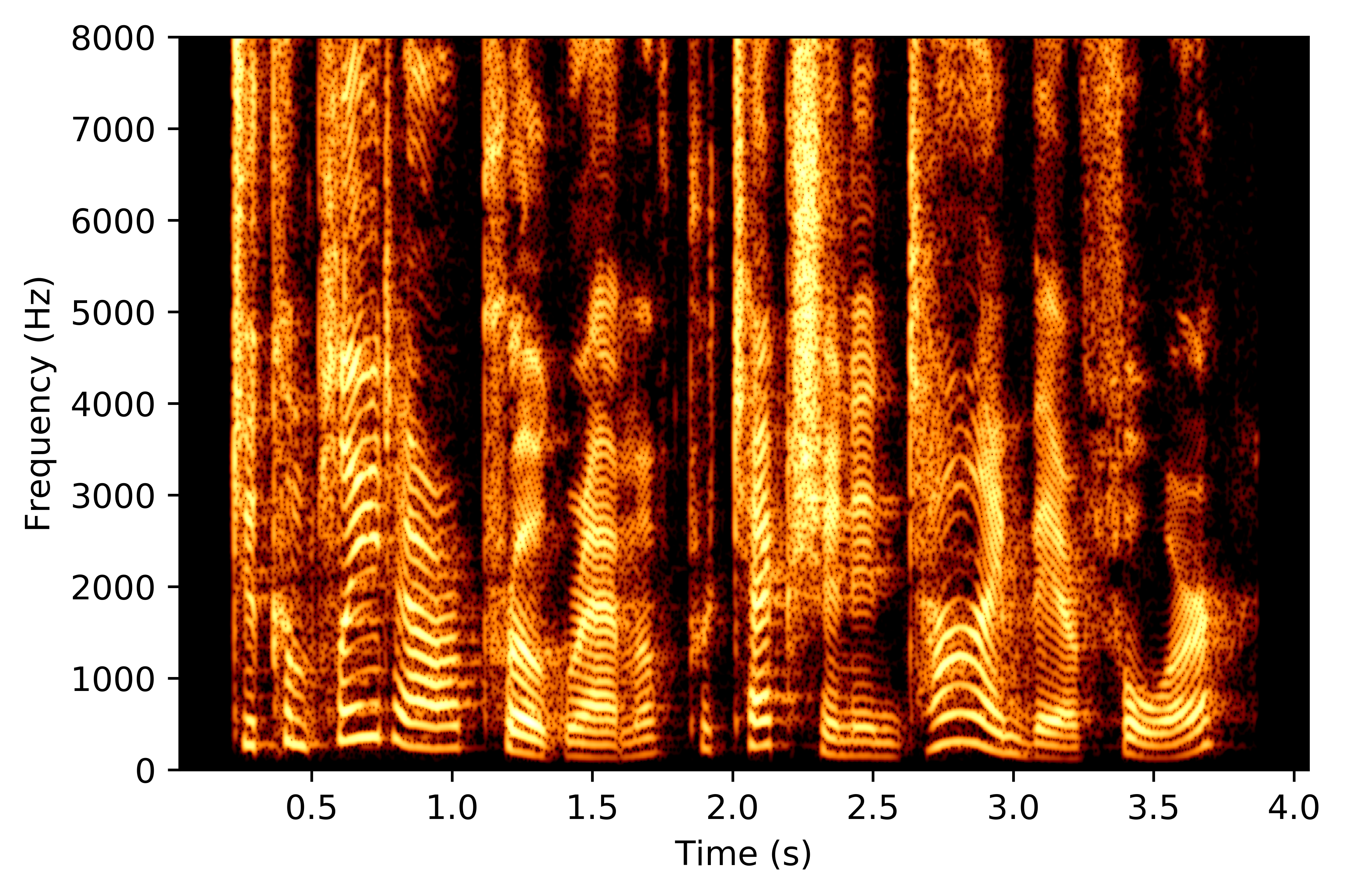}}
\subfigure[Reference audio]{\label{fig:5b}\includegraphics[width=0.31\textwidth]{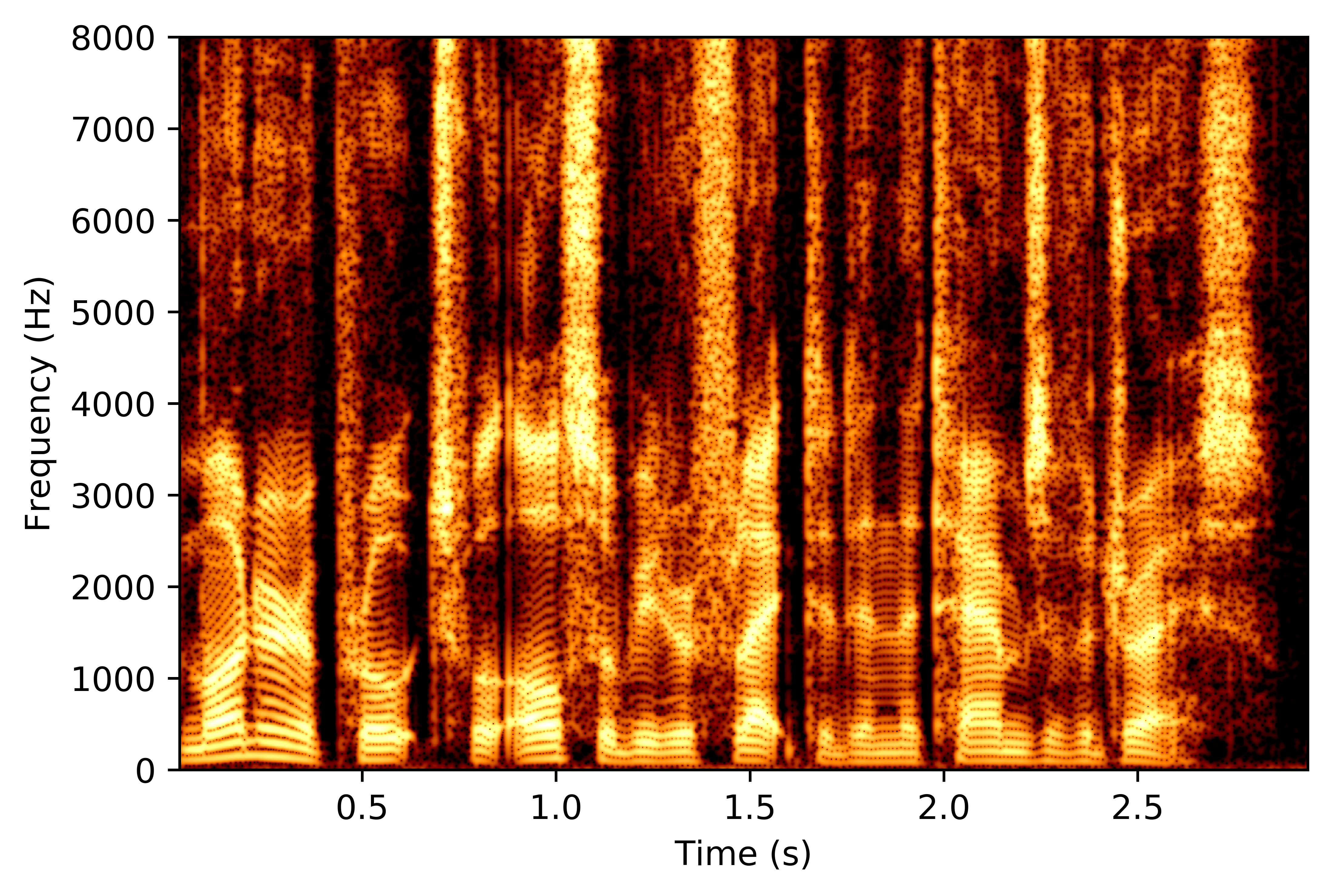}}
\subfigure[Converted audio]{\label{fig:5c}\includegraphics[width=0.31\textwidth]{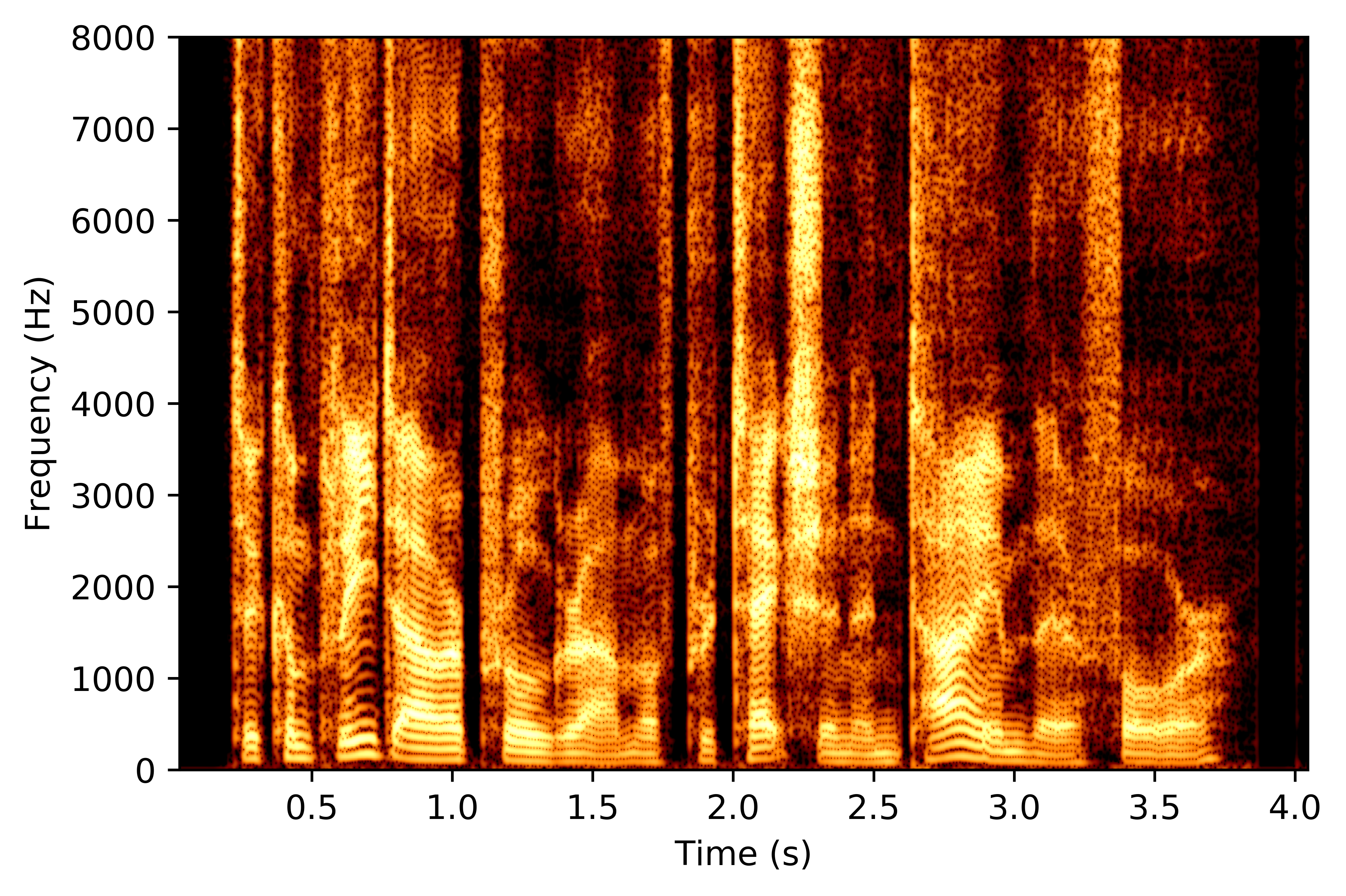}}
\caption{An example of any-to-any voice conversion. The source audio is from the LJSpeech dataset and the reference audio is from the VCTK dataset, both unseen during training. }
\label{fig:5}
\end{figure*}

Since our text encoder, text aligner, pitch extractor, and decoder are trained in a speaker-agnostic manner, our decoder can reconstruct speech from any aligned phonemes, pitch, energy, and reference speakers. Therefore, our model can perform any-to-any voice conversion by extracting the alignment, pitch, and energy from an input mel-spectrogram and generating speech using a style vector of reference audio from an arbitrary target speaker. Our voice conversion scheme is transcription-guided, similar to Mellotron \cite{valle2020mellotron} and Cotatron \cite{park2020cotatron}.  We provide one example in Figure \ref{fig:5} with both source and target speaker unseen from the LJSpeech and VCTK datasets. We refer our readers to our demo page for more examples.

\section{Detailed Model Architectures}
\label{app:C}

In this section, we provide detailed model architectures of StyleTTS, which consists of eight modules. Since we use the same text encoder as in Tacorton 2 \cite{shen2018natural}, very similar architecture to the decoder of Tacotron 2 for text aligner and the same architecture as the JDC network \cite{kum2019joint} for pitch extractor, we leave the readers to the above references for detailed descriptions of these modules. Here, we only provide detailed architectures for the other five modules. All activation functions used are leaky ReLU (LReLU) with a negative slope of 0.2. We apply spectral normalization \cite{miyato2018spectral} to all trainable parameters in style encoder and discriminator and weight normalization \cite{salimans2016weight} to those in decoder because they are shown to be beneficial for adversarial training. 

\textbf{Decoder } (Table \ref{tab:8}). Our decoder takes four inputs: the aligned phoneme representation, the pitch F0, the energy, and the style code. It consists of seven 1-D residual blocks (ResBlk) along with three sub-modules for processing the input F0, energy, and residual of the phoneme representation. The normalization consists of both instance normalization (IN) and adaptive instance normalization (AdaIN). We concatenate (Concat) the processed F0, energy, and residual of phonemes with the output from each residual block as the input to the next block for the first three blocks.


\begin{table*}[!h]
\small
\caption{Style encoder and discriminator architectures. $T$ represents the input length of the mel-spectrogram, and $D$ is the output dimension. For style encoder, $D=128$. For discriminator, $D = 1$. }
\label{tab:9}
\centering
\begin{tabular}{cccc}
\hline
Layer    & Pooling      & Norm & Output Shape \\ \hline

Mel $\bm{x}$                     & -                           & -                        & 1$\times$80$\times T$                           \\ \hline
Conv 1$\times$1                  & -                           & -                        & 64$\times$80$\times T$                          \\
ResBlk                    & Dilated Conv                & -                        & 128$\times$40$\times T/2   $                    \\
ResBlk                    & Dilated Conv                & -                        & 256$\times$20$\times T/4$                       \\
ResBlk                    & Dilated Conv                & -                        & 512$\times$10$\times T/8$                       \\
ResBlk                    & Dilated Conv                & -                        & 512$\times$5$\times T/16 $                      \\ \hline
LReLU                     & -                           & -                        & 512$\times$5$\times T/16$                       \\
Conv 5$\times$5                  & -                           & -                        & 512$\times$1$\times T/80 $                      \\
LReLU                     & -                           & -                        & 512$\times$1$\times T/80$                       \\ \hline
-                         & AdaAvg                      & -                        & 512$\times$1                            \\
Linear                    & -                           & -                        & $D \times$1                            \\ \hline
\end{tabular}
\end{table*}

\begin{table*}
\small
\caption{Duration and prosody predictor architectures. $N$ represents the number of input phonemes and $T$ represents the length of the alignment. $h_\text{text}$ is the hidden phoneme representation from the text encoder, $d_\text{align}$ is the monotonic alignment with shape $N \times T$, $\bm{s}$ is the style code, $a_\text{pred}$ is the predicted duration, $p_\text{pred}$ is the predicted pitch and $\norm{x}_\text{pred}$ is the predicted energy. $h_\text{prosody}$ and $h_\text{aprosody}$ are intermediate outputs from submodules. }
\label{tab:12}
\centering
\begin{tabular}{cccccc}
\hline
Submodule                            & External Input       & Layer   & Norm  & Output Shape  & \makecell{Submodule \\Output}             \\ \hline
\multirow{6}{*}{Prosody Encoder}     & $h_\text{text}$, $\bm{s}$                 & Concat  & -     & (512 + 128)$\times N$ & \multirow{6}{*}{ $h_\text{prosody}$}  \\
                                     & $\bm{s}$                    & bi-LSTM & AdaLN & 512$\times N$         &                              \\
                                     & $\bm{s}$                    & Concat  & -     & (512 + 128)$\times N$ &                              \\
                                     & $\bm{s}$                    & bi-LSTM & AdaLN & 512$\times N$         &                              \\
                                     & $\bm{s}$                    & Concat  & -     & (512 + 128)$\times N$ &                              \\
                                     & $\bm{s}$                    & bi-LSTM & AdaLN & 512$\times N$         &                              \\ \hline
\multirow{2}{*}{Duration Projection} & $h_\text{prosody}$           & bi-LSTM & -     & 512$\times N$         & \multirow{2}{*}{$a_\text{pred}$}     \\
                                     & -                    & Linear  & -     & 1$\times N$           &                              \\ \hline
\multirow{3}{*}{Shared LSTM}         & $h_\text{prosody}$, $d_\text{align}$ & Dot     & -     & 512$\times T$         & \multirow{3}{*}{$h_\text{aprosody}$} \\
                                     & $\bm{s}$                    & Concat  & -     & (512 + 128)$\times T$ &                              \\
                                     & -                    & bi-LSTM & -     & 512$\times T$         &                              \\ \hline
\multirow{4}{*}{Pitch Predictor}     & $h_\text{aprosody}$, $\bm{s}$                    & ResBlk  & AdaIN & 512$\times T$         & \multirow{4}{*}{$p_\text{pred}$}     \\
                                     & $\bm{s}$                    & ResBlk  & AdaIN & 256$\times T$         &                              \\
                                     & $\bm{s}$                    & ResBlk  & AdaIN & 256$\times T$         &                              \\
                                     & -                    & Linear  & -     & 1$\times T$           &                              \\ \hline
\multirow{4}{*}{Energy Predictor}    & $h_\text{aprosody}$, $\bm{s}$                    & ResBlk  & AdaIN & 512$\times T$         & \multirow{4}{*}{$\norm{x}_\text{pred}$}     \\
                                     & $\bm{s}$                    & ResBlk  & AdaIN & 256$\times T$         &                              \\
                                     & $\bm{s}$                    & ResBlk  & AdaIN & 256$\times T$         &                              \\
                                     & -                    & Linear  & - & 1$\times T$           &                              \\ \hline
\end{tabular}
\end{table*}

\textbf{Style Encoder and Discriminator} (Table \ref{tab:9}). Our style encoder and discriminator share the same architecture, which consists of four 2-D residual blocks (ResBlk). The dimension of the style vector is set to 128. We use learned weights for pooling through a dilated convolution (Dilated Conv) layer with a kernel size of 3$\times$3. We apply an adaptive average pooling (AdaAvg) along the time axis of the feature map to make the output independent of the size of the input mel-spectrogram.

\textbf{Duration and Prosody Predictors} (Table \ref{tab:12}). The duration predictor and prosody predictors are trained together in the second stage of training. There is a shared 3-layer bidirectional LSTM (bi-LSTM)  $\bm{s}$ between the duration predictor and prosody predictor named text feature encoder, each followed by an adaptive layer normalization (AdaLN). AdaLN is similar to AdaIN where the gain and bias are predicted from the style vector $\bm{s}$. However, unlike AdaIN which normalizes each channel independently, AdaLN normalizes the entire feature map. The style vector $\bm{s}$ is also concatenated (Concat) with the output to every token from each LSTM layer as the input to the next LSTM layer. Lastly, we have a final bidirectional LSTM and a linear projection $L$ that maps $h_\text{prosody}$ into the predicted duration.

The hidden representation $h_\text{prosody}$ is dotted with the alignment $d_\text{align}$ and sent to the prosody decoder. The prosody encoder consists of one bidirectional LSTM and two sets of three residual blocks (ResBlk) with AdaIN followed by a linear projection, one for predicting the F0 and another for predicting the energy, respectively. \\

\end{appendices}

\end{document}